\documentclass[aps,prc,twocolumn,showpacs,preprintnumbers,nofootinbib,float,superscriptaddress, amsmath,amssymb,aps,longbibliography]{revtex4-2}

\usepackage{enumitem}

\usepackage{graphicx}
\usepackage{dcolumn}
\usepackage{bm}

\usepackage{hyperref}
\hypersetup{
colorlinks=true,
linkcolor=blue,
citecolor=blue,
urlcolor=blue
}

\usepackage{lineno}
\newcommand*\patchAmsMathEnvironmentForLineno[1]{%
\expandafter\let\csname old#1\expandafter\endcsname\csname #1\endcsname
\expandafter\let\csname oldend#1\expandafter\endcsname\csname end#1\endcsname
\renewenvironment{#1}%
{\linenomath\csname old#1\endcsname}%
{\csname oldend#1\endcsname\endlinenomath}}%
\newcommand*\patchBothAmsMathEnvironmentsForLineno[1]{%
\patchAmsMathEnvironmentForLineno{#1}%
\patchAmsMathEnvironmentForLineno{#1*}}%
\AtBeginDocument{%
\patchBothAmsMathEnvironmentsForLineno{equation}%
\patchBothAmsMathEnvironmentsForLineno{align}%
\patchBothAmsMathEnvironmentsForLineno{flalign}%
\patchBothAmsMathEnvironmentsForLineno{alignat}%
\patchBothAmsMathEnvironmentsForLineno{gather}%
\patchBothAmsMathEnvironmentsForLineno{multline}%
}

\usepackage{ulem}
\usepackage{xcolor}
\DeclareRobustCommand{\erase}{\bgroup\markoverwith{\textcolor{red}{\rule[.5ex]{2pt}{1.5pt}}}\ULon}

\makeatletter
\newsavebox{\@brx}
\newcommand{\llangle}[1][]{\savebox{\@brx}{\(\m@th{#1\langle}\)}%
  \mathopen{\copy\@brx\mkern2mu\kern-0.9\wd\@brx\usebox{\@brx}}}
\newcommand{\rrangle}[1][]{\savebox{\@brx}{\(\m@th{#1\rangle}\)}%
  \mathclose{\copy\@brx\mkern2mu\kern-0.9\wd\@brx\usebox{\@brx}}}
\makeatother

\begin{document}


\title{Rapidity-dependent charge dynamics in high-energy nuclear collisions\\from dynamical core-corona initialization}

\author{Shin-ei Fujii}
\altaffiliation{Present affiliation: Zhejiang Institute of Modern Physics, Department of Physics, Zhejiang University, Hangzhou 310027, China}
\email{s-fujii-8c2@eagle.sophia.ac.jp}
\affiliation{Department of Physics, Sophia University, Tokyo 102-8554, Japan}

\author{Yasuki Tachibana}
\altaffiliation{Present affiliation: Department of Space Systems Engineering, Kyushu Institute of Technology, Kitakyushu, 804-8550, Japan}
\email{tachibana.yasuki935@mail.kyutech.jp}
\affiliation{Akita International University, Yuwa, Akita-city 010-1292, Japan}
\affiliation{Department of Physics, Sophia University, Tokyo 102-8554, Japan}

\author{Tetsufumi Hirano}%
 \email{hirano@sophia.ac.jp}
\affiliation{Department of Physics, Sophia University, Tokyo 102-8554, Japan}

\date{\today}

\begin{abstract} 
We develop a new model that provides a unified description of the equilibration and subsequent transport of conserved charges, including baryon number, electric charge, and strangeness, as well as the dynamical formation and evolution of the quark-gluon plasma medium in high-energy nuclear collisions.
This is achieved by extending the state-of-the-art model of dynamical core-corona initialization to incorporate phenomenological source terms for conserved charges within a full (3+1)-dimensional hydrodynamic framework; the updated model is referred to as DCCI3. 
Consequently, the model simultaneously describes the core (equilibrium) and corona (nonequilibrium), including their mutual interactions.
We present a detailed construction of the DCCI3 model and employ it in numerical simulations of Pb+Pb collisions at $\sqrt{s_{NN}}=2.76~\mathrm{TeV}$ to investigate the resulting dynamics of matter and conserved charges, with particular emphasis on the longitudinal structure. 
We find that each of the chemical potentials exhibits sizable fluctuations around midrapidity, which challenges the conventional assumption of vanishing values as an oversimplification. 
Moreover, the baryon chemical potential increases sharply with rapidity and becomes comparable to the temperature at forward rapidities.
This suggests that the Rapidity Scan (RS) approach in high-energy nuclear collisions can serve as a complementary probe to the Beam Energy Scan (BES) in exploring the QCD phase diagram. 

\end{abstract}

\maketitle


\section{INTRODUCTION}\label{sec:intro}
Quantum chromodynamics (QCD) predicts a rich phase structure of strongly interacting matter as a function of temperature and baryon chemical potential.
At sufficiently high temperature or density, hadronic matter undergoes a transition to a deconfined state known as the quark-gluon plasma (QGP). 
High-energy nuclear collisions at Relativistic Heavy Ion Collider (RHIC) and the Large Hadron Collider (LHC) provide a unique opportunity to create and study such matter under extreme conditions. 

When the matter created in high-energy nuclear collisions approaches local equilibrium, its subsequent space-time evolution can be successfully described by relativistic hydrodynamics.
Indeed, early ideal hydrodynamic models that assumed rapid thermalization and nearly perfect fluid behavior achieved remarkable success in reproducing bulk observables around midrapidity~\cite{Heinz:2001xi, Kolb:2000fha, Teaney:2000cw, Hirano:2001eu, Hirano:2002ds, Hirano:2005wx, Lee:2005gw}.
In modern modeling, dynamical frameworks that consistently incorporate all stages of the collision\textemdash{}including the pre-equilibrium dynamics, hydrodynamic evolution, and subsequent hadronic transport\textemdash{}have been developed, enabling a more detailed and quantitative understanding of the collision dynamics (see, e.g., Refs.~\cite{Hirano:2012kj, Putschke:2019yrg, Schenke:2020mbo, An:2021wof, Kanakubo:2021qcw}).

While modern dynamical frameworks have achieved remarkable success in describing the energy--momentum evolution of the system, the role of conserved charges has been comparatively under-explored. 
At high collision energies, the colliding nuclei are highly Lorentz contracted, where the Lorentz factor $\gamma\sim 1000$, leading to an extremely short interaction time. 
This allows the incident nuclei, which carry the charges, to pass through with minimal stopping, sweeping the charges toward forward rapidities. 
Consequently, the conserved charge densities and the associated chemical potentials are often assumed to be zero throughout the system in conventional studies, largely because of their focus on the midrapidity region. 
Although these simplifications have facilitated the development of dynamical models, realistic high-energy nuclear collisions inherently involve event-by-event fluctuations of conserved charges. 
Even though the net baryon number, electric charge, and strangeness are negligible 
at midrapidity, local fluctuations can render these quantities finite on an 
event-by-event basis~\cite{Carzon:2019qja, Plumberg:2024leb, Gardim:2024nyz}. 
Such fluctuations have been shown to induce non-negligible effects on final-state observables, such as particle-antiparticle asymmetries in identified-particle flow~\cite{Gardim:2024nyz}.

At forward rapidity, the role of conserved charges becomes even more pronounced. A substantial fraction of the baryon number is carried by leading baryons\footnote{In this context, the term \textit{leading baryon} denotes baryons near the beam rapidity, rather than those leading in transverse momentum.}, and several studies suggest that the forward rapidity region may reach extremely high baryon densities through nuclear compression~\cite{Gyulassy:1986fk, Li:2018yvx, Li:2018ini}.
Such a baryon-rich environment can leave characteristic imprints on observables; for instance, the suppression of thermal dilepton production has been proposed as a possible probe of high-density QCD matter at forward rapidity~\cite{Oya:2026too}. 
These perspectives highlight the forward rapidity region as a novel environment 
for accessing finite-density QCD matter even at high collision energies and also 
open a new frontier for exploring high-density QCD matter beyond conventional 
approaches such as the Beam Energy Scan (BES) program at RHIC~\cite{Mohanty:2011nm, Schmah:2013vea, Odyniec:2013aaa, McDonald:2015tza, Tlusty:2018rif}. 
In addition to the theoretical interest, several upcoming experimental programs, such as the ALICE FoCal detector upgrade~\cite{ALICE:2023rol, Aehle:2023mpy, ALICE:2024jtt, Huhta:2024xpm, Inaba:2025boq} and the proposed ALICE3 project~\cite{Adamova:2019vkf, ALICE:2022wwr}, will provide extended coverage including the forward rapidity region.
To enable meaningful discussions in this region, it is essential to first develop a detailed understanding of the matter produced there. Such an understanding will play a foundational role in the anticipated expansion of studies in the so-called forward physics. 

To quantitatively describe conserved charge dynamics in high-energy nuclear collisions, a dynamical framework that consistently incorporates both equilibration and subsequent evolution of conserved charges is required.
In particular, it is essential to start from a realistic description of non-equilibrium matter production, in which initial charge fluctuations naturally arise, and to follow their evolution through interactions with the medium toward local equilibration.
Furthermore, the degree of equilibration is not yet known in the forward rapidity region.
A realistic model must therefore be capable of describing QGP formation, charge equilibration, and transport, as well as their interplay with non-equilibrium components across the entire system.
This calls for a fully three-dimensional dynamical framework that can consistently treat the system from midrapidity to forward rapidity.

The dynamical core–corona initialization (DCCI) framework~\cite{Kanakubo:2018vkl, Kanakubo:2019ogh, Kanakubo:2021qcw, Kanakubo:2022ual, Kanakubo:2022jju, Ito:2026hec} provides a unified description of high-energy nuclear collisions by incorporating the core-corona picture, in which nonequilibrium partons gradually generate locally equilibrated matter through interactions.
This framework has successfully reproduced a wide range of observables, spanning low to high $p_T$ and small to large collision systems.
In the present study, we extend the DCCI2~\cite{Kanakubo:2019ogh, Kanakubo:2021qcw, Kanakubo:2022ual, Kanakubo:2022jju, Ito:2026hec} model to include the dynamics of conserved charges and develop a new framework, referred to as DCCI3.
In DCCI3, not only energy–momentum but also baryon number ($B$), electric charge ($Q$), and strangeness ($S$) are dynamically deposited into the hydrodynamic medium in a fully conserved manner, which naturally introduces event-by-event fluctuations of conserved charge densities.
The dynamical source terms further enable us to quantify the degree of equilibration even in the forward rapidity region and to evaluate the corresponding chemical potentials.
The subsequent evolution is described within a $(3+1)$-dimensional hydrodynamic framework with a four-dimensional equation of state~\cite{Monnai:2024pvy}.
For consistency, the particlization procedure is generalized to finite chemical potentials ($\mu_{{B}}$, $\mu_{{Q}}$, $\mu_{{S}}$), allowing for a unified treatment of conserved-charge dynamics throughout the entire evolution.

We apply DCCI3 to Pb+Pb collisions at $\sqrt{s_{{NN}}} = 2.76\,\mathrm{TeV}$ and systematically analyze the dynamical evolution of conserved charges and their associated chemical potentials across the entire system, from midrapidity to forward rapidity.
In particular, we aim to elucidate the event-by-event fluctuations and the longitudinal structure 
arising from conserved charge dynamics, and to assess their impact on the final-state observables.
In Sec.~\ref{sec:model}, we describe the detailed construction of 
the DCCI3 model, particularly focusing on the implementation of the 
source terms, fluidization dynamics, and the color string 
modification consistent with charge conservation. 
In Sec.~\ref{sec:results}, we present the results obtained from the DCCI3 model, with primary emphasis on the fluctuations and longitudinal structures of conserved charges. 
We also discuss net hadron production and the distribution of conserved charges carried by the final-state hadrons.
Section~\ref{sec:summary} is devoted to the summary.
Building on the framework and charge dynamics established in this paper, we investigate their phenomenological consequences for experimentally accessible hadron observables, with particular emphasis on rapidity-dependent equilibration and strange hadron production, in a forthcoming paper.

Throughout this paper, we use natural units, $\hbar = c = k_{B} = 1$, and the mostly minus convention of the metric, $g_{\mu\nu} = \mathrm{diag}(+,-,-,-)$.

\section{MODEL}\label{sec:model}
\begin{figure}[htbp]
\includegraphics[width=\linewidth]{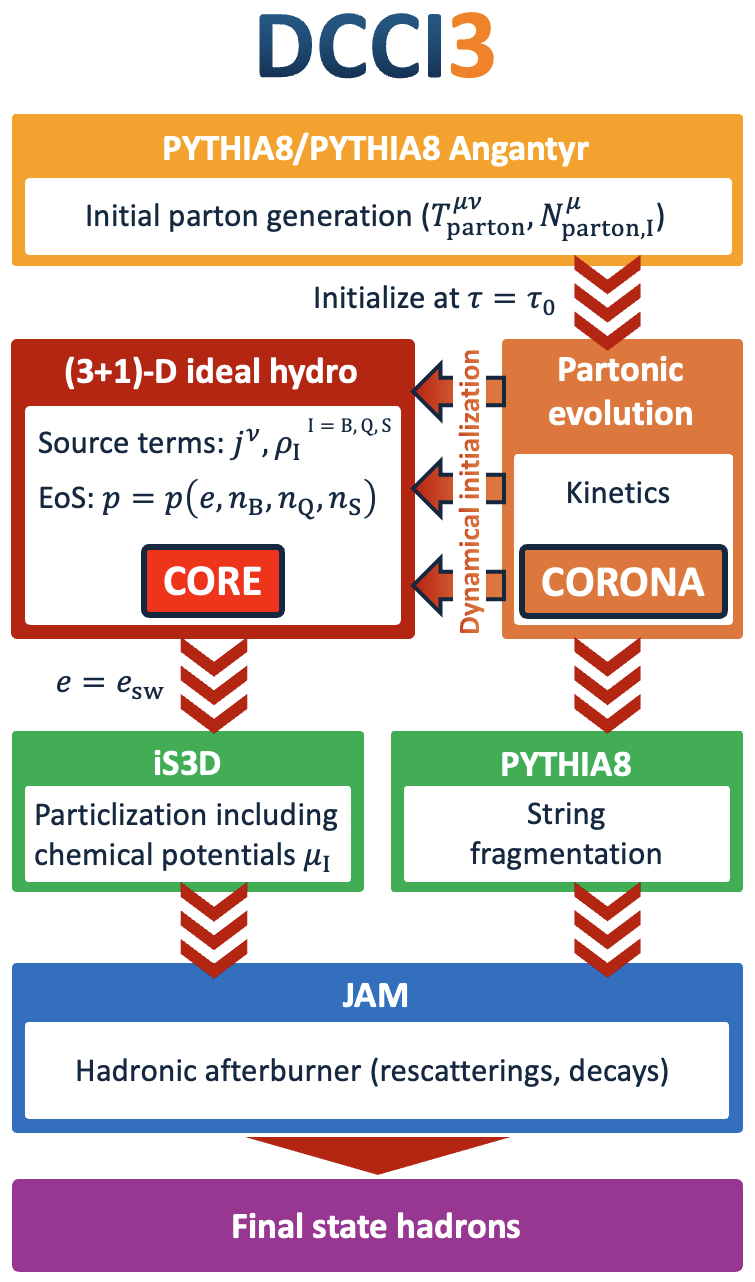}
\caption{Workflow of the DCCI3 model. Details are shown in the text.}
\label{fig:DCCI3}
\end{figure}

The DCCI3 model is an integrated dynamical framework in which each stage of the collision reaction is described by an appropriate theoretical approach, enabling a consistent description from the initial state to final-state hadron production. 
The overall structure of the model is summarized in Fig.~\ref{fig:DCCI3}.

The simulation begins with initial parton production using PYTHIA8~\cite{Sjostrand:2007gs, Bierlich:2014xba}, utilizing the Angantyr model~\cite{Bierlich:2016smv, Bierlich:2017sxk, Bierlich:2018xfw} specifically for heavy-ion projectiles. 
Initially, no locally equilibrated medium is present. As the produced partons propagate in spacetime, they deposit their energy, momentum, and conserved charges, dynamically building up a fluid. In practice, this process is implemented by introducing source terms in the hydrodynamic equations, whose strength depends on the local parton density and their momenta.
Within this framework, the resulting thermalized fluid is referred to as the \textit{core}, while the partons that do not fully thermalize and survive as nonequilibrium particles are designated as the \textit{corona}.

The spacetime evolution of the core is described by (3+1)-dimensional ideal hydrodynamics using the four-dimensional equation of state, NEOS-4D~\cite{Monnai:2024pvy}, which incorporates degrees of freedom for baryon number, electric charge, and strangeness. The transition to a particle-based description occurs at a predefined switching energy density ($e_{\mathrm{sw}}$) via the Cooper–Frye prescription~\cite{Cooper:1974mv}, extended to finite chemical potentials using the iS3D particlization module~\cite{McNelis:2019auj}. 
In parallel, non-equilibrated partons remaining as the corona are hadronized through string fragmentation implemented in PYTHIA8.
Finally, all produced hadrons from both the core and the corona are fed into a hadronic cascade model JAM~\cite{Nara:1999dz} for the subsequent hadronic rescatterings and decays. 

In the following sections, we describe each component of DCCI3, 
which extends DCCI2 by incorporating conserved charge dynamics. 
For details on DCCI2, we refer the reader to Ref.~\cite{Kanakubo:2021qcw}. 

\subsection{Initial partons}\label{sec:initial_parton}
In the DCCI3 framework, initial partons are generated by using the general-purpose event generator PYTHIA8~\cite{Sjostrand:2007gs, Bierlich:2014xba}. 
When heavy-ion projectiles are specified, PYTHIA invokes the Angantyr model~\cite{Bierlich:2016smv, Bierlich:2017sxk, Bierlich:2018xfw}, describing the collision as a superposition of multiple nucleon--nucleon sub-collisions, generating an ensemble of excited strings. 
The use of PYTHIA8, including its Angantyr extension for heavy-ion collisions, ensures that the model starts from a realistic microscopic description of parton production by incorporating multiple parton interactions and nuclear geometry. 
In the present framework, we directly use these partons prior to hadronization, obtained 
by disabling hadronization through the setting \texttt{HadronLevel:all=off} in PYTHIA8. 
These partons serve as the input to the subsequent dynamical core-corona initialization process, and act as localized sources for the formation of the QGP fluid. 

\subsection{Dynamical core-corona initialization}\label{sec:DCCI}
In DCCI3, all partons generated by PYTHIA8 subsequently undergo the dynamical core–corona initialization process, which is initiated globally at a proper time of $\tau_0 = 0.1~\mathrm{fm}$. At the onset of this process, the system consists solely of nonequilibrium partons. The fluid component, referred to as the core, is then dynamically built up through energy--momentum deposition from these partons as they propagate, while conserved charges carried by the partons, such as baryon number, electric charge, and strangeness, are incorporated into the fluid. It should be noted that the energy--momentum deposition occurs continuously and depends on the local parton density and the kinematics of each parton. As a result, some partons may not fully fluidize and instead survive as nonequilibrium degrees of freedom, forming the corona. 


\subsubsection{Hydrodynamic equations with source terms}
In the dynamical core-corona initialization process, the system consists of two components: nonequilibrium partons described within a particle-based framework and QGP fluid components that have reached local equilibrium. 
For the entire system, the total energy, momentum, and the conserved charges---baryon number, electric charge, and strangeness, additionally incorporated in DCCI3---must satisfy the conservation laws, yielding the following continuity equations: 
\begin{align}
\partial_\mu T^{\mu\nu}_{\mathrm{total}}(x) &= \partial_\mu \left[ T^{\mu\nu}_{\mathrm{fluid}}(x) + T^{\mu\nu}_{\mathrm{parton}}(x) \right] = 0,\label{eq:conservation_energy}\\
\partial_\mu N^{\mu}_{I, \mathrm{total}}(x) &= \partial_\mu \left[ N^{\mu}_{I, \mathrm{fluid}}(x) + N^{\mu}_{I,\mathrm{parton}}(x) \right] = 0,\label{eq:conservation_charge}
\end{align}
where $T^{\mu\nu}_{\mathrm{fluid}}$, $T^{\mu\nu}_{\mathrm{parton}}$, and $T^{\mu\nu}_{\mathrm{total}}$ denote the energy--momentum tensors of the QGP fluid, the nonequilibrium partons, and the entire system, respectively, and $N^{\mu}_{I,\mathrm{fluid}}$, $N^{\mu}_{I,\mathrm{parton}}$, and $N^{\mu}_{I,\mathrm{total}}$ are the corresponding conserved charge currents.
The subscript $I$ on the charge currents denotes $I = B, Q$, and $S$. 

In the current DCCI3 framework, the fluid is treated as an ideal relativistic fluid. Accordingly, the energy--momentum tensor $T^{\mu\nu}_\mathrm{fluid}$ and the conserved charge currents $N^{\mu}_{\mathrm{fluid},I}$ are given by the constitutive relations: 
\begin{align}
T_{\mathrm{fluid}}^{\mu\nu} &= (e + P)u^\mu u^\nu - Pg^{\mu\nu},\\
N^{\mu}_{I,\mathrm{fluid}} &= n_{I} u^{\mu},
\end{align}
where $e$, $P$, $n_I$, and $u^\mu$ denote the energy density, pressure, conserved charge densities, and fluid four-velocity, respectively. 
The thermodynamic quantities $e$, $P$, and $n_I$ ($I = B, Q, S$) are related through the equation of state $P = P(e, n_{B}, n_{Q}, n_{S})$, for which we employ the four-dimensional equation of state NEOS-4D~\cite{Monnai:2024pvy}. 
Then Eqs.~\eqref{eq:conservation_energy}~and~\eqref{eq:conservation_charge} lead, for the fluid component, to the following hydrodynamic equations with source terms that govern its space-time evolution:
\begin{align}
\partial_\mu T^{\mu\nu}_{\mathrm{fluid}}(x) &= j^\nu (x),\\
\partial_\mu N^{\mu}_{I,\mathrm{fluid}}(x) &= \rho_{I}(x),\label{eq:charge_eom}
\end{align}
where the source terms are determined from the partonic components as 
\begin{align}
j^\nu (x) &= -\partial_\mu T^{\mu\nu}_{\mathrm{parton}}(x)\\
\rho_{I}(x) &= -\partial_\mu N^{\mu}_{I,\mathrm{parton}}(x).
\end{align}
Thus, any energy--momentum and conserved charges lost by the partons are gained by the fluid. 


In high-energy nuclear collisions, the medium undergoes a strongly anisotropic expansion, especially along the longitudinal (beam) direction. In such cases, Milne coordinates $(\tau, x, y, \eta_{s})$, where $\tau = \sqrt{t^2 - z^2}$ is the proper time and $\eta_{s} = (1/2)\ln [(t+z)/(t-z)]$ is the space-time rapidity, are commonly used, as they allow for much more efficient numerical simulations of the longitudinal dynamics. While the hydrodynamic equations were written above in Cartesian coordinates, the numerical calculations in the DCCI framework are performed in Milne coordinates. The equations actually solved are the extended form of the hydrodynamic equations, in which the ordinary partial derivatives are replaced by covariant derivatives ($\nabla_\mu$) appropriate for Milne coordinates: 
\begin{align}
\nabla_\mu T^{\mu\nu}_{\mathrm{fluid}}(x) &= j^\nu (x),\\
\nabla_\mu N^{\mu}_{I,\mathrm{fluid}}(x) &= \rho_{I}(x).\label{eq:charge_eom}
\end{align}
In the following, all expressions are given in Milne coordinates, and the indices of vectors and tensors are taken in this coordinate system unless otherwise specified. The time evolution of the system is described in terms of the proper time $\tau$.

The initial condition for the fluid component is set such that no fluid is present at the initial time $\tau=\tau_0$, i.e., $T^{\mu\nu}_{\mathrm{fluid}}(\tau=\tau_0)=0$ and $N^{\mu}_{I,\mathrm{fluid}}(\tau=\tau_0)=0$. The fluid is then dynamically built up through the continuous deposition of energy–momentum via the source term $j^\nu$, which serves as the key element coupling the parton dynamics to the hydrodynamic fields, while the conserved charges are supplied through $\rho_I$.

\subsubsection{Source terms}
Under the assumption that the energy–momentum deposited by partons undergoes instantaneous thermalization and immediately forms a locally equilibrated fluid, 
the explicit form of the source term $j^{\nu}(x)$ can be constructed from the microscopic expression of the partonic energy--momentum tensor $T^{\mu\nu}_{\mathrm{parton}}$ as
\begin{align}
j^{\nu}(x) = \sum_i \left(-\frac{d p^{\nu}_i(\tau)}{d\tau}\right) G_{\sigma} \left(\bm{x}-\bm{x}_i(\tau)\right),
\label{eq:source_term}
\end{align}
where $-dp_i^\nu/d\tau$ is the four-momentum loss rate of parton $i$, in other words, the fluidization rate per particle in the dynamical core-corona picture. 
Meanwhile, $G_{\sigma}$ in Eq.~\eqref{eq:source_term} specifies the spatial profile of the source associated with each parton; here, it is modeled as a Gaussian distribution with widths $\sigma$ centered at the position of the $i$th parton, $\bm{x}_i(\tau) = (x_i(\tau),y_i(\tau),\eta_{s,i}(\tau))$: 
\begin{align}
G_{\sigma} \left(\bm{x}-\bm{x}_i(\tau)\right)&= \frac{1}{2\pi \sigma_\perp^2} \exp\left[ -\frac{(\bm{x}_{\perp} - \bm{x}_{\perp,i}(\tau))^2}{2\sigma_\perp^2} \right] \nonumber\\
&\hspace{1em}\times \frac{1}{\sqrt{2\pi \sigma_{\eta_s}^2 \tau^2}} \exp\left[ -\frac{(\eta_{s} - \eta_{s,i}(\tau))^2}{2\sigma_{\eta_s}^2} \right],\label{eq:gaussian_density}
\end{align}
where $\bm{x}_{\perp} = (x,y)$. The parameters $\sigma_\perp$ and $\sigma_{\eta_s}$ denote the Gaussian widths in the transverse ($x$ and $y$) and longitudinal ($\eta_s$) directions, respectively, and are treated as free parameters in DCCI3.


In analogy to the energy--momentum source term, we define the conserved charge source terms $\rho_{I}(x)$ as a sum over contributions from partons that carry charge $I$ and are depositing it into the fluid:
\begin{equation}
\rho_{I} (x) = \sum_{j}^{n_\mathrm{dead}} \left(-\frac{dI_{j}(\tau)}{d\tau} \right) G_{\sigma} \left( \bm{x} - \bm{x}_j(\tau) \right), \label{eq:charge_source}
\end{equation}
where $I_{j}$ is the quantum charge $I\, (=B,~Q,~S)$ carried by the $j$th parton, which is completely absorbed into the fluid. 
Such absorbed partons are referred to as dead partons in DCCI3; further details are provided in Secs.~\ref{sec:color_string} and~\ref{sec:deposition_scenario}. 
Accordingly, $-dI_{j}/d\tau$ represents the rate of charge transfer from the partons to the fluid, which contributes to the local fluid charge density $n_{I}(x)$ via the source term. 
Here, it should be noted that the Gaussian width of the profile in the conserved charge source terms $\rho_{I}(x)$ of Eq.~\eqref{eq:charge_source} could, in principle, differ from that in the energy–momentum source term $j^\nu(x)$ of Eq.~\eqref{eq:source_term}, but we set them equal for simplicity.

\subsubsection{Fluidization rate}
The conversion of parton energy and momentum into the QGP fluid is driven by microscopic interactions among partons.
Partons that undergo frequent secondary scatterings are expected to thermalize and contribute to the formation of the QGP fluid, whereas those experiencing only a few interactions tend to retain their energy–momentum and remain non-fluidized. 
To describe this behavior, the fluidization rate $-dp_i^\nu/d\tau$ is modeled phenomenologically by introducing an effective drag force in the parton equation of motion, whose magnitude depends on the local density of surrounding particles and their relative velocity: 
\begin{equation}
-\frac{dp^{\mu}_i}{d\tau} = \sum_{j}^{n_\mathrm{coll}} \sigma_{ij} \rho_{ij} |v_{\text{rel},ij}| p^{\mu}_i. \label{eq:fluidization_rate} 
\end{equation}
The sum runs over all partons $j$ interacting with the nonequilibrium $i$th parton, including both nonequilibrium and thermally sampled equilibrium partons, thereby accounting for both corona–corona and corona–core interactions. Details of the thermal parton sampling procedure are provided in Sec.~\ref{sec:thermal_parton_sampling}. 
The $\sigma_{ij}$ is an effective cross-section between parton $i$ and parton $j$, $\rho_{ij}$ is the effective density of parton $j$ seen by parton $i$, and $|v_{\text{rel},ij}|$ is the magnitude of the relative velocity between the $i$th parton and the $j$th parton. 
Since the mean free path $l_{ij}$ between $i$ and $j$ can be estimated as $l_{ij} = 1/\sigma_{ij}\rho_{ij}$, the combination $\sigma_{ij}\rho_{ij}|v_{\text{rel},ij}|$ can be regarded as scattering rate per unit time for parton $i$. 
Equation~\eqref{eq:fluidization_rate} implies that the fractional momentum loss per unit time is proportional to the scattering rate, leading to a substantial reduction of the four-momentum when the $i$th parton propagates in a dense medium. Moreover, the trajectory of the propagating parton is implicitly assumed to be straight, i.e., no deflection is considered (eikonal approximation), while it gradually loses energy.

In the DCCI3 framework, the cross-section $\sigma_{ij}$ is modeled as a function of collision energy of partons $s_{ij} = ( p^{\mu}_i + p^{\mu}_j)^2$ to capture the core-corona picture in the momentum space:
\begin{equation}
\sigma_{ij} = \text{min} \left\{ \frac{\sigma_0}{s_{ij}/s_0}, \pi b_{\rm cut}^2 \right\},\label{eq:sigma}
\end{equation}
where $\sigma_0$ is an adjustable parameter, $s_0 = 1~\mathrm{GeV}^2$ is a reference scale for normalization, and $b_{\rm cut}$ is introduced to avoid infrared divergence for small $s_{ij}$. 
Due to the collision energy dependence of the cross-section $\sigma_{ij}$, partons with significantly higher energies are less likely to deposit their energy and momentum during the evolution.
Using this cross-section $\sigma_{ij}$, a collision between partons $i$ and $j$ is considered to occur if their spatial distance is smaller than $\sqrt{\sigma_{ij}/\pi}$. 
The effective density profile $\rho_{ij}$ is assumed to have the same functional form as the source profile given in Eq.~\eqref{eq:gaussian_density}, but with different widths denoted by $w_\perp$ and $w_{\eta_s}$: 
\begin{align}
\rho_{ij} = G_{w}\left(\bm{x}_i(\tau)-\bm{x}_j(\tau)\right).
\label{eq:rho}
\end{align}

Regarding the conserved charge fluidization rate $-dI_{j}/d\tau$, the subscript $j$ only accounts for dead partons: partons that have lost their entire energy and momentum into the fluid and are no longer active in subsequent evolution.
This treatment is in contrast to the energy--momentum source term in Eq.\eqref{eq:source_term} discussed earlier, where all traversing partons contribute continuously as they propagate through the fluid. 
This distinction arises from a fundamental difference between the nature of the conserved quantities: energy and momentum are continuous, whereas conserved charges such as baryon number, electric charge, and strangeness are intrinsically discrete. Treating these charges as being deposited gradually would lead to unphysical fractional charge propagation, since they cannot be meaningfully divided or partially transferred. Accordingly, a parton transfers its conserved charges to the fluid only upon being fully absorbed.
Thus, the fluidization rate of conserved charges, $-dI_{j}/d\tau$, is given by 
\begin{equation}
-\frac{dI_{j}}{d\tau} = I_j \delta(\tau-\tau_{\mathrm{dead},j}),\label{eq:charge_source_delta}
\end{equation}
where $\tau_{\mathrm{dead},j}$ denotes the proper time at which parton $j$ has completely deposited its energy and momentum into the fluid and becomes an inactive ``dead'' parton. 
By the end of the evolution, the total conserved charges carried by both the string-fragmented partons and the fluidized partons are guaranteed to remain appropriately discretized quantum numbers separately. 

To summarize the mechanism: each parton acts as a moving source term for hydrodynamics and gradually ``melts'' into the fluid, depending on the local environment, such as the surrounding parton density, their momenta, and the interaction strength. 
Through this continuous process, energy and momentum are deposited into the medium, while conserved charges are transferred to the fluid in a manner consistent with their discrete nature, leading to the dynamical formation of the core fluid without introducing any sharp or discontinuous separation between the core and corona.
Moreover, the dynamical core-corona initialization does not involve any explicit rapidity dependence, thereby providing a natural and universal description of fluid formation with conserved charges consistently incorporated along the longitudinal direction. 
This is an important feature for investigating our primary interest in this work: fluid formation in the forward region and the baryon density within the resulting medium. 


\subsubsection{Thermal parton sampling}\label{sec:thermal_parton_sampling}
To evaluate energy--momentum loss of Eq.~\eqref{eq:fluidization_rate} for each nonequilibrium parton traversing the core, one needs to sample partons from the QGP fluid. 
At each time step, under the massless ideal-gas approximation, the parton number density in the QGP fluid is estimated as 
\begin{equation}
n = \frac{90 d’ \zeta(3)}{4\pi^4 d} s(e, n_{B}, n_{Q}, n_{S}). 
\label{eq:number_density}
\end{equation} 
Here, $s=s(e, n_{B}, n_{Q}, n_{S})$ is the local entropy density calculated from the NEOS-4D equation of state, and $d$ and $d'$ are degrees of freedom defined as 
\begin{equation}
d = d_F \times \frac{7}{8} + d_B, \quad d' = d_F \times \frac{3}{4} + d_B,
\end{equation}
where $d_F = 36$ and $d_B = 16$ are degrees of freedom of fermions and bosons, assuming three quark flavors (u, d, s). \footnote{Although the degrees of freedom are assumed to be u, d, or s, the thermal partons are used only for the evaluation of energy--momentum loss rate. As a result, the quark flavor does not play a role, and no flavor information is assigned to the partons.}
The number of partons in each fluid cell is then sampled from a Poisson distribution based on the local number density $n$, and their species—quark, antiquark, or gluon—are assigned probabilistically according to their statistical degrees of freedom.

Each sampled parton is further assigned a three-momentum $\bm{k}$ according to the corresponding normalized massless thermal distribution, i.e., the Bose or Fermi distribution. The momentum is first generated in the local rest frame of the fluid cell and then boosted to the laboratory frame via a Lorentz transformation using the local flow velocity field $u^\mu$ of the fluid. The spatial positions of the sampled partons are uniformly distributed within each fluid cell.

\subsubsection{Equation of state}
To close the system of the hydrodynamic equations, we employ a four-dimensional equation of state, NEOS-4D~\cite{Monnai:2024pvy}, in which baryon number, electric charge, and strangeness are treated as independent thermodynamic degrees of freedom. 
The NEOS model incorporates susceptibilities obtained from lattice QCD simulations at vanishing densities~\cite{Borsanyi:2011sw, Bellwied:2015lba, Borsanyi:2018grb, HotQCD:2012fhj, Ding:2015fca, Bazavov:2017dus} to construct a finite-density EoS using the Taylor expansion method~\cite{Gavai:2001fr, Allton:2002zi} at high temperatures. In the high-temperature QGP phase, only light quarks, u, d, and s, are considered.
At lower temperatures, where the fugacity expansion becomes unreliable~\cite{Karsch:2010hm}, NEOS adopts the hadron resonance gas (HRG) model, where the hadron list is restricted to those composed of u, d, and s valence quarks. 
This hybrid construction ensures a smooth transition between the QGP and hadronic phases with finite chemical potentials. 
In DCCI3, the two-dimensional equations of state, NEOS-BQS (-BS, -B)~\cite{Monnai:2019hkn, Monnai:2021kgu}, in which electric charge and strangeness are subject to specific constraints, can also be employed as alternative options. See, Appendix~\ref{sec:NEOSBQS} for results from NEOS-BQS.
This enables direct comparison with the fully unconstrained NEOS-4D and allows us to isolate and assess the impact of fluctuations in $\mu_{Q}$ and $\mu_{S}$.

\subsection{Modification of color strings}\label{sec:color_string}
After the dynamical core-corona initialization, partons that survive without losing all their energy are hadronized, as described in Sec.~\ref{sec:hadronization_corona}, using the Lund string model of PYTHIA8. 
In this procedure, the color information originally assigned to these partons at their production stage in PYTHIA8 is employed. For hadronization to proceed, the parton system must form an overall color-neutral configuration; however, this condition is generally violated during the dynamical core-corona initialization.
When some partons deposit all their energy into the fluid and become dead partons, the color charges they carried are removed from the system, leaving an incomplete color structure. 
In DCCI3, the color string structure is modified to maintain overall color neutrality throughout the space-time evolution. To this end, string connections are dynamically adjusted in response to interactions with the medium and the emergence of dead partons, as detailed below.

\subsubsection{String modification at $\tau=\tau_s$}
In DCCI3, a specific string modification time $\tau = \tau_s\,(>\tau_0)$ is introduced, beyond which the modification of color strings is initiated. 
As the first step of this procedure, at $\tau = \tau_s$, the modification is performed based on the spatial relationship between the string-connected partons and the regions of hot, deconfined QGP medium characterized by an energy density $e > e_{\mathrm{sw}}$, as well as on the invariant mass of each string. According to these considerations, the color strings are modified following predefined decision criteria. 
\begin{figure}[htbp]
    \centering
    \includegraphics[clip,width=0.95\linewidth]{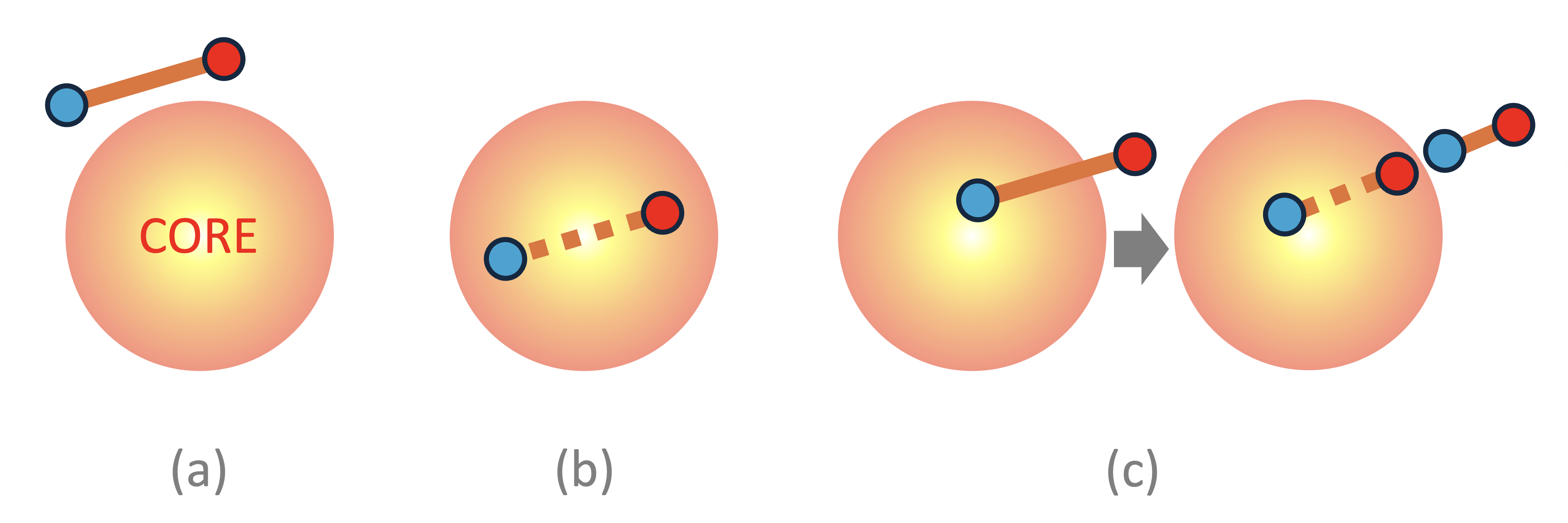}
    \caption[Three different scenarios of string]
    {Three scenarios of strings are illustrated: (a) a string entirely outside the hot core medium, (b) a string embedded in the hot core medium, and (c) a string crossing the hot core boundary. Small red and blue circles represent quarks and antiquarks, respectively, while orange lines between them denote strings. Strings inside the medium are melted and depicted as dotted orange lines. The large light-red circles indicate the hot core medium with energy denisity $e > e_{\mathrm{sw}}$.}
    \label{fig:Cutting}
\end{figure}
\begin{itemize}
\item[(a)] String entirely outside the hot core: 
As shown in Fig.~\ref{fig:Cutting} (a), if a string's constituents lie entirely outside the hot medium—defined by an energy density $e > e_{\mathrm{sw}}$, where $e_{\mathrm{sw}}$ is a chosen switching energy density indicating the core boundary—at $\tau_s$, then its color string configuration is left unchanged. It subsequently hadronizes via the Lund string fragmentation model as in vacuum, without undergoing further energy loss or contributing to additional fluid formation.


\item[(b)] String embedded in the hot core:
If a string is fully immersed in the hot medium as illustrated in Fig.~\ref{fig:Cutting} (b), i.e., the entire chain of connected partons is surrounded by fluid with $e > e_{\mathrm{sw}}$, the string tension can no longer be sustained, preventing its fragmentation. Under these conditions, the string is effectively ``melted,'' and its color connections are dissolved, consistent with the absence of confinement in the deconfined phase.
The constituent quarks and gluons are then treated as individual nonequilibrium partons propagating through the medium (referred to as \textit{surviving partons}), which continue to undergo partonic interactions and continuously deposit energy and momentum into the fluid. 
While some are eventually absorbed into the fluid (becoming dead partons), others escape from the hot medium while retaining finite energy and hadronize at later times.

\item[(c)] String crossing the hot core boundary: The most interesting case is when a string straddles the boundary of the hot medium region, defined by the iso-energy-density surface at $e = e_{\mathrm{sw}}$, with part of it inside and part outside, as illustrated in Fig.~\ref{fig:Cutting} (c).
Because a color string physically cannot continuously extend into a deconfined hot medium, the string is cut at the interface. 
At the severed ends, a thermal quark--antiquark pair is sampled from the medium and attached to the truncated string endpoints. This generates shortened, color-neutral strings on both the medium and vacuum sides. Consequently, the portion outside the hot medium is converted into a shorter string that immediately hadronizes via string fragmentation. Meanwhile, the medium-side portion is treated as a string embedded in the hot core discussed above, behaving as surviving or dead partons.
\end{itemize}

\subsubsection{Hard strings}
High-$p_T$ partons are expected to interact more weakly with the medium than soft components, and their fragmentation pattern is therefore not significantly modified. This is reflected in the transverse momentum spectra of high-$p_T$ hadrons, which do not exhibit strong modifications. Motivated by this consideration, a criterion to identify strings associated with high-$p_T$ partons is introduced. Specifically, a transverse momentum cut is imposed, and any string containing at least one parton with $p_T > p_{T,\text{cut}}$ is classified as a \textit{hard string}. For such hard strings, no modification of the string structure is applied, even when they cross the core boundary or are fully embedded in the medium.

\subsubsection{Ministrings}
In addition to spatial criteria, the invariant mass of each color-neutral string is monitored to decide if it should fragment.
In PYTHIA8, hadronization via the Lund string fragmentation algorithm requires the invariant mass to exceed a minimum threshold.
If a string's invariant mass falls below a specified value $m_{\text{th}}$, for example, due to very low energy or substantial shortening, the model does not hadronize the string through the standard fragmentation procedure~\cite{Norrbin:2000zc}.
Such low-mass strings (referred to as \textit{ministrings}) are instead treated differently in DCCI3: their constituent partons are immediately absorbed into the fluid, becoming dead partons.
This prevents trying to fragment a string that does not have enough energy to produce even the lightest hadrons.
This threshold check is applied both at the initial $\tau_s$ evaluation and later during any string modification process.
In DCCI3, the invariant mass threshold is set to be same as the PYTHIA8 default setting~\cite{Norrbin:2000zc}:
\begin{equation}
m_{\text{th}} = m_1 + m_2 + 1.0~\mathrm{GeV},
\end{equation}
where $m_1$ and $m_2$ are the masses of partons at the edge of the string.

\subsubsection{Escaping partons from the fluid}
At $\tau > \tau_s$, nonequilibrium partons that were kept inside 
the fluid\textemdash{}the surviving partons that did not fragment 
at $\tau_s$\textemdash{}can later hadronize if they escape from the 
QGP fluid before they become completely absorbed as dead partons. 
These individual partons are monitored as they propagate through the dense and hot medium.
When the surviving parton moves into a region below the switching energy density, $e < e_{\mathrm{sw}}$, it is considered to have exited the QGP and reached the core boundary.
At the moment of exit, a parton pairing occurs analogous to a string cutting as illustrated in Fig.~\ref{fig:Cutting} (c). 
For example, imagine a quark that has traversed the QGP and is now streaming out; as it emerges, it grabs a thermal antiquark from the fluid cell at the crossing point so that together they form a color-neutral quark--antiquark pair, which undergoes the string fragmentation routine to produce hadrons.
This prescription is exactly the concept of the coalescence models~\cite{Hwa:2003ic, Hwa:2004ng, Greco:2003xt, Isse:2007pa, Fries:2008hs, Han:2016uhh, Fries:2019vws, Zhao:2020wcd, Kordella:2020nwi}, which is mainly responsible for generating intermediate $p_T$ hadrons.

\subsubsection{Parton sampling at the boundary}
\begin{figure}[htbp]
    \centering
    \includegraphics[clip,width=0.95\linewidth]{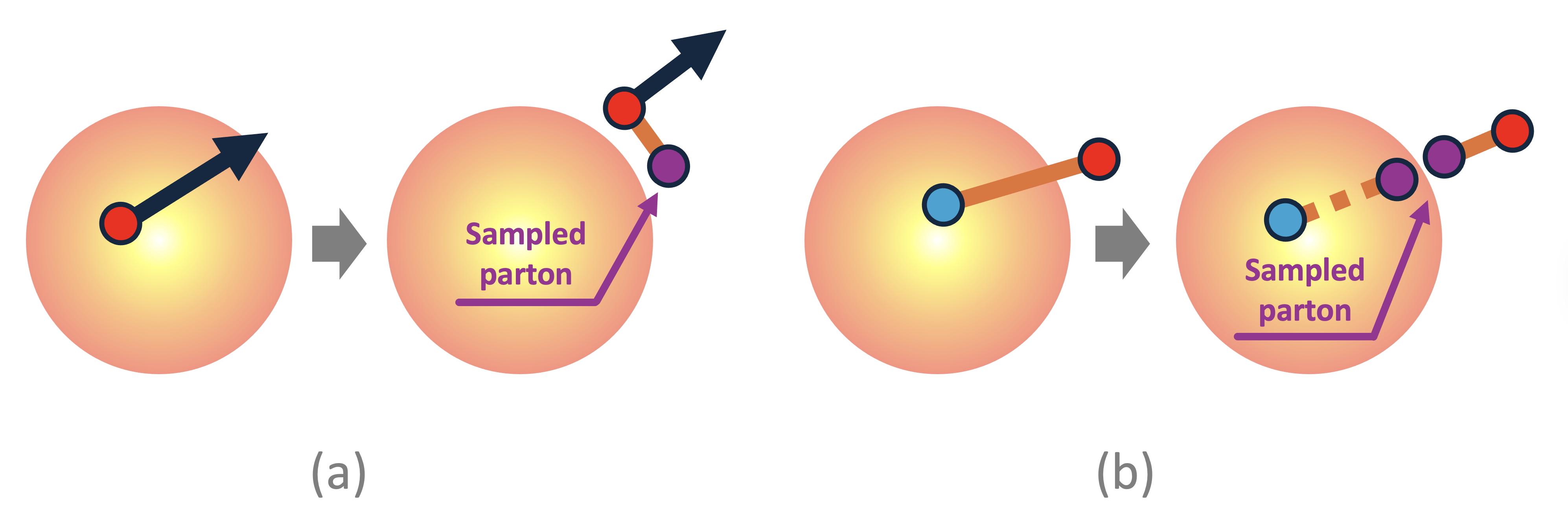}
    \caption[Two different scenarios of paprton sampling]{
    Two different scenarios of parton sampling from the medium are illustrated: (a) sampling associated with an escaping parton from the medium and (b) sampling associated with string cutting at the core boundary. The purple circles represent the sampled partons.}
    \label{fig:Sampling}
\end{figure}
Whenever a surviving parton escapes from the medium [Fig.~\ref{fig:Sampling} (a)] or a string is cut at the core boundary [Fig.~\ref{fig:Sampling} (b)], partons are sampled to pair with it.
This ensures color conservation, forming a color-neutral object, and allows hadronization to proceed for that pair.

Regarding the parton pairing associated with the escaping parton as illustrated in Fig.~\ref{fig:Sampling} (a), a parton is sampled from the thermal distribution.
The flavor of the sampled parton, light quarks (u, d, s) or gluons, is chosen stochastically based on the weights calculated from the thermal distribution function, including chemical potentials.
The assignment of momentum is identical to that used in the thermal parton sampling explained in Sec.~\ref{sec:thermal_parton_sampling}, i.e., it gives the newly sampled parton a momentum typical of the thermal environment.
It is important to note that the energy--momentum corresponding to a sampled new parton is not subtracted from the fluid, leading to a slight energy--momentum violation in the system at $\tau \geq \tau_s$.
The energy--momentum in the entire system is exactly conserved at $\tau_0 \leq \tau < \tau_s$, since no string modification is done before $\tau_s$.
In previous implementations in DCCI2~\cite{Kanakubo:2021qcw, Kanakubo:2022jju}, thermal distributions without chemical potentials had been employed, but in the present work, the effects of chemical potentials are incorporated so that the relevance of the flavor composition is preserved.

When the string is cut at the boundary, the parton sampling mechanism should differ from that of parton sampling associated with escaping partons, since the process involves the formation of a $q$--$\bar{q}$ pair. 
Accordingly, the sampling should be based on the vacuum pair-production probability rather than a thermal distribution [Fig.~\ref{fig:Sampling} (b)]. 
In this work, the flavor ratios are fixed to the default \textsc{Pythia}8 setting, $\mathrm{u}:\mathrm{d}:\mathrm{s}=1:1:0.217$.
Furthermore, the flavor of the sampled parton is strictly constrained to match that of its partner, i.e., if a quark of flavor $f(\mathrm{u},\mathrm{d},\mathrm{s})$ is sampled from the boundary, it is paired with an antiquark of the same flavor $\bar{f}(\mathrm{\bar{u}},\mathrm{\bar{d}},\mathrm{\bar{s}})$, and vice versa.
This ensures that all three conserved charges, $B$, $Q$, and $S$, are preserved locally at the pair creation level, not just in the statistical average.
This extension is essential in the DCCI3 framework, where the conservation of net charges is treated dynamically.
Without this constraint, unphysical fluctuations in electric charge or strangeness could be introduced during the string modification process.
By enforcing flavor-consistent pairing, we guarantee that the global conservation of $B$,~$Q$,~and~$S$ is maintained even in the interface between the core and the corona.
However, the momenta of the sampled partons are still drawn from thermal distributions, and this aspect should also be revised to a string-based prescription in future work.\\

After all surviving partons are processed in these ways explained above, the only matter remaining is fluid component, which is then converted to hadrons via the usual Cooper--Frye particlization~\cite{Cooper:1974mv}, which is handled by iS3D~\cite{McNelis:2019auj} in the DCCI3 framework. 

\subsection{Deposition scenarios of conserved charges}\label{sec:deposition_scenario}
To maintain the conservation law of charges even during the complicated string modification process at $\tau_s$ and subsequent evolution, four distinct scenarios of charge deposition are introduced. 
These scenarios are referred to as dead parton, coalescence, string cutting, and re-pairing.
Each of these scenarios is explained in detail below. 

\begin{figure}[htbp]
    \centering
    \includegraphics[clip,width=0.95\linewidth]{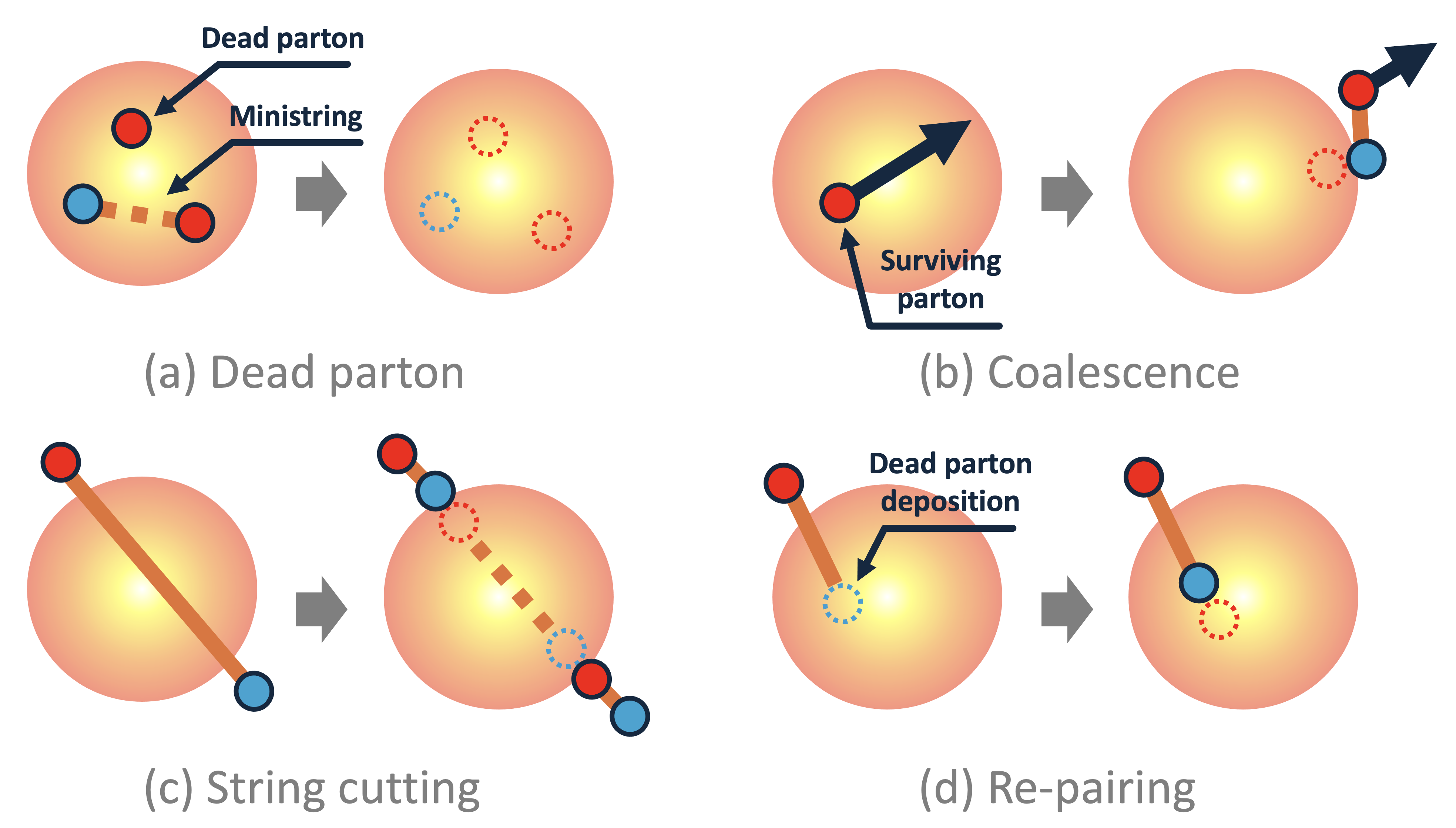}
    \caption[Four different scenarios of deposition]{
    Four scenarios of conserved charge deposition are illustrated: (a) Dead parton, (b) Coalescence, (c) String cutting, and (d) Re-pairing. Dotted red and dotted blue circles denote charge deposition of quarks and antiquarks, respectively.
    }
    \label{fig:deposition_scenario}
\end{figure}

\begin{itemize}
\item[(a)] Dead parton:
Figure~\ref{fig:deposition_scenario} (a) illustrates the simplest charge deposition scenario, referred to as the dead parton. As explained in Sec.~\ref{sec:color_string}, when partons lose all of their energy to the fluid or when the invariant mass of the string falls below the threshold $m_{\mathrm{th}}$, such partons are completely absorbed into the fluid and are removed from the nonequilibrium parton list. 
At the moment of absorption, the charges carried by the dead partons are deposited into the fluid. This process constitutes the primary mechanism of charge equilibration in the present framework.

\item[(b)] Coalescence:
The second scenario, coalescence, is shown in Fig.~\ref{fig:deposition_scenario} (b). As already depicted in Fig.~\ref{fig:Sampling} (a), when a surviving parton is escaping from the medium, a thermal parton is picked up from the medium to form a color-neutral string. Since the pickup of a parton also carries its associated charges, in order to satisfy charge conservation in the whole system, the charges of the corresponding antiparticle are deposited into the fluid. 

\item[(c)] String cutting:
The third scenario, string cutting, is illustrated in Fig.~\ref{fig:deposition_scenario} (c). 
Depending on the geometric configuration between the string and the hot medium boundary, the string cutting procedure can produce new strings that are attached only to sampled thermal partons. In such cases, these strings are regarded as unphysical and are therefore discarded together with the sampled thermal partons. To maintain charge conservation during this string cutting process, the corresponding charges are deposited into the fluid. 

\item[(d)] Re-pairing:
The final scenario, re-pairing, is illustrated in Fig.~\ref{fig:deposition_scenario} (d). When a string crosses the fluid boundary and a parton on the medium side loses all of its energy, charge deposition proceeds according to the scenario (a). Although all partons should remain connected by a string at least until $\tau_s$ even if they are embedded in the medium, this situation can occur just before the string modification at $\tau_s$. To restore the string, a parton is sampled from the medium, and immediately afterward, string modification is performed. Associated with this thermal parton sampling, the corresponding antiparticle's charges are deposited.

\end{itemize}
By incorporating all these deposition scenarios into the DCCI model, baryon number, electric charge, and strangeness are strictly conserved in the entire system at any stage of the evolution before particlization.

Before closing this section, let us mention the time ordering of the energy--momentum source term [Eq.~(\ref{eq:source_term})] and the conserved charge source term [Eq.~(\ref{eq:charge_source})].
The energy--momentum deposition and the subsequent formation of the fluid start at $\tau_0$, whereas the deposition of conserved charges is activated only after the starting time of string modification $\tau_s$ described in Sec.~\ref{sec:color_string}.

\subsection{Particlization and hadronic afterburner}
\subsubsection{Particlization of the core}
To perform particlization from the core fluid for event-by-event simulations and to provide input to the hadronic afterburner, the open-source code, iS3D~\cite{McNelis:2019auj}, which samples hadrons from the continuous Cooper–Frye spectrum~\cite{Cooper:1974mv} using a Monte Carlo procedure, is employed in the DCCI3 framework. 
The switching hypersurface is constructed as an iso-energy-density surface with $e_{\mathrm{sw}} = 0.547\,\mathrm{GeV/fm^3}$, which approximately corresponds to a temperature of $T \approx 165\,\mathrm{MeV}$ in the limit $n_{I}\rightarrow 0$ within the NEOS-4D equation of state. 

While the original iS3D code optionally allows for the inclusion of the baryon chemical potential $\mu_{B}$ in the equilibrium distribution function, the electric charge and strangeness chemical potentials, $\mu_{Q}$ and $\mu_{S}$, are not included.
In the present work, we extend iS3D to incorporate the full set of chemical potentials $(\mu_{B},\,\mu_{Q},\,\mu_{S})$, enabling a consistent implementation of BQS charge dynamics in the particlization procedure.
In addition, the default hadron list in iS3D is replaced with that used in the hadron cascade model JAM~\cite{Nara:1999dz}, which serves as the hadronic afterburner, thereby ensuring consistency throughout the entire framework.

\subsubsection{Hadronization of the corona}
\label{sec:hadronization_corona}
Surviving partons that escape the medium carry the modified color string information described in Sec.~\ref{sec:color_string}, and are subsequently hadronized using the Lund string fragmentation model implemented in PYTHIA8. 
Thus, the framework provides a unified hybrid description that combines Cooper-Frye particlization for the core with Lund string fragmentation for the corona, thereby enabling a consistent treatment across the full range of event multiplicities, from vacuum-like fragmentation in low-multiplicity events to thermal particle production in high-multiplicity events.

\subsubsection{Hadronic afterburner}
DCCI3 simulates the hadronic afterburner for the late-stage evolution by passing all produced hadrons to a hadron cascade model JAM (Jet-AA Microscopic Transport Model)~\cite{Nara:1999dz}. 
To ensure consistency, unstable particles from iS3D and PYTHIA8 are transferred without prior decay; JAM then manages their full space-time evolution, including rescatterings and decays.\footnote{Note that electroweak decays, except $\Sigma^0 \rightarrow \Lambda + \gamma$, are turned off to facilitate direct comparison with experimental observables.}
This approach naturally enables interactions between core and corona hadrons while ensuring consistent resonance feed-down. Furthermore, JAM allows rescattering to be toggled on or off, facilitating the isolation of final-state interaction effects.


\subsection{Parameter set in DCCI3}
Simulations employing the DCCI3 model require the specification of several parameters detailed in this section. 
Most of these parameters are inherited from the previous version of the model, DCCI2, having been tuned to reproduce observables sensitive to the core–corona composition of particle production. 
In particular, the parameters are constrained by the $\mathrm{\Omega/\pi}$ yield ratios measured by the ALICE Collaboration~\cite{ALICE:2013xmt, ALICE:2016fzo}, which exhibit a strong sensitivity to the relative contributions of the core and corona components. At the same time, the parameters are chosen to provide a reasonable description of the charged-particle multiplicity at midrapidity as a function of event multiplicity or collision centrality.

Table~\ref{tb:parameters} summarizes the key parameters used in the DCCI3 framework alongside their assigned values. The parameter \texttt{pT0Ref} in PYTHIA8 governs the multi-parton interactions (MPI) during the initial parton generation (see, Ref.~\cite{Kanakubo:2021qcw} for details). 
The time-step sizes $\Delta\tau_1$ and $\Delta\tau_2$ are applied to the stages before and after $\tau_s$, respectively; this distinction is made because the initial fluidization stage $(\tau_0 \le \tau \le \tau_s)$ involves rapid dynamics, whereas the subsequent hydrodynamic evolution $(\tau > \tau_s)$ proceeds more gradually.
Finally, $\Delta x$, $\Delta y$, and $\Delta \eta_s$ denote the grid spacings employed in the hydrodynamic simulation. 


\begin{table}[htbp]
\begin{ruledtabular}
\begin{tabular}{ccl}
\textbf{Parameter} & \textbf{Value} & \textbf{Description}\\
\hline
$p_{T0\mathrm{Ref}}$                & 1.0 GeV & parameter in PYTHIA\\
$e_{\mathrm{sw}}$                   & $0.547~\mathrm{GeV/fm^3}$ & switching energy density\\
$\sigma_0$                          & $0.3~\rm{fm^2}$ & cross-section\\
$b_{\mathrm{cut}}$                  & 1.0 fm & cutting parameter\\
$p_{T,\mathrm{cut}}$                & 3.0 GeV & criteria of high $p_T$ parton\\
$w_\perp$                           & 0.5 fm & width of density profile in $x$--$y$\\
$w_{\eta_s}$                        & 0.5 & width of density profile in $\eta_s$\\
$\sigma_\perp$                      & 0.6 fm & width of deposition in $x$--$y$\\
$\sigma_{\eta_s}$                   & 0.5 & width of deposition in $\eta_s$\\
$\tau_0$                            & 0.1 fm & initialization time\\
$\tau_s$                            & 0.3 fm & string modification time\\
$\Delta\tau_1$                      & 0.01 fm & discrete time in $\tau_0 \leq \tau \leq \tau_s$\\
$\Delta\tau_2$                      & 0.3 fm & discrete time in $\tau_s < \tau$\\
$\Delta x$                          & 0.3 fm & grid spacing in $x$\\
$\Delta y$                          & 0.3 fm & grid spacing in $y$\\
$\Delta \eta_s$                     & 0.15 & grid spacing in $\eta_s$\\
\end{tabular}
\end{ruledtabular}
\caption[List of model parameters used in the present study]{List of model parameters used in the present study.}
\label{tb:parameters}
\end{table}

\section{RESULTS AND DISCUSSIONS}\label{sec:results}
In this study, we focus on Pb+Pb collisions at $\sqrt{s_{{NN}}} = 2.76\,\mathrm{TeV}$, as our primary interest lies in conserved charge dynamics in high-energy heavy-ion collisions at the LHC.
When presenting results, the notations $\langle \cdots \rangle$ and $\llangle \cdots \rrangle$ are used. 
The former denotes an event-averaged quantity, while the latter denotes a quantity averaged over both events and hypersurface elements.


\subsection{Time evolution of energy and conserved charges}\label{sec:conservation}
\begin{figure*}[htbp]
\includegraphics[width=0.95\linewidth]{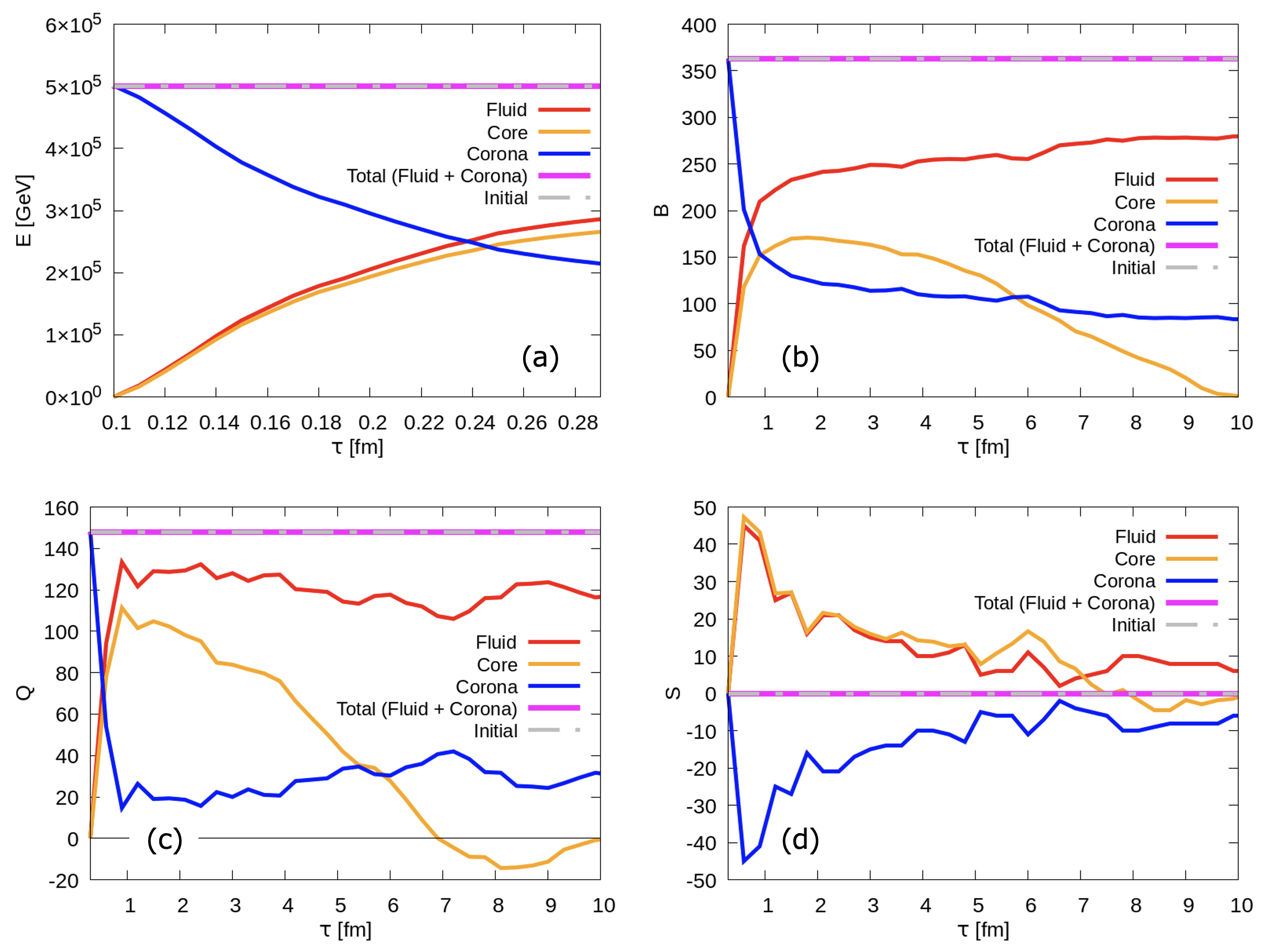}
\caption{Time evolution of energy and conserved charges in a randomly selected central event of Pb+Pb collisions at $\sqrt{s_{NN}} = 2.76\,\mathrm{TeV}$. In this event, the total energy and charges are $E \approx 5.01 \times 10^{5}\, \mathrm{GeV}$, $B=363$, $Q=148$, and $S=0$, respectively. Please note that the spectators are excluded here. Panels show (a) energy ($\tau_0 \leq \tau < \tau_s$), (b) baryon number ($\tau_s \leq \tau$), (c) electric charge ($\tau_s \leq \tau$), and (d) strangeness ($\tau_s \leq \tau$). The contributions from all fluid (red), core (fluid with $e \geq e_{\mathrm{sw}}$; orange), corona (surviving partons + fragmented strings; blue), and total (red + blue; magenta) are shown. This total remains, at all times, identical to the energy and conserved charges of the initial parton ensemble generated by PYTHIA8 (dash-dotted gray).}
\label{fig:conservation}
\end{figure*}
As it is evident from Eqs.~(\ref{eq:conservation_energy}) and~(\ref{eq:conservation_charge}), energy, momentum, and three charges are conserved as a whole system.
Within the DCCI3 framework, the sum of the core and the corona is therefore always conserved, while energy, momentum, and charges can be exchanged between them. 
At each timestep, the total energy and conserved charges are separated into four contributions:
\begin{enumerate}[label={(\roman*)}]
\item contributions from surviving nonequilibrium partons of the corona, \label{parton-corona}
\item contributions from fragmented strings, representing the already-hadronized corona, \label{hadron-corona}
\item contributions from the hot medium core fluid with $e \geq e_{\mathrm{sw}}$, and 
\label{active-core}
\item contributions from the already-particlized core, having reached $e < e_{\mathrm{sw}}$. \label{particlized-core}
\end{enumerate} 
By construction, the sum of these contributions reproduces the total energy and conserved charges determined from the initial parton list generated by PYTHIA8.

Figure~\ref{fig:conservation} shows the time evolution of energy and conserved charges obtained from a single, randomly chosen central event. 
The solid red lines denote the total fluidized component at a given time, which encompasses both the hot core before particlization~\ref{active-core} and the amount of core already particlized~\ref{particlized-core}. 
Similarly, the solid blue lines represent the total corona contribution, consisting of surviving partons still depositing energy and momentum into the fluid~\ref{parton-corona}, as well as the already-hadronized corona~\ref{hadron-corona}. 
In all panels, the sum of the total fluidized component and the corona contribution (solid magenta) remains constant throughout the evolution and perfectly matches the initial value (dash-dotted gray), demonstrating exact conservation of energy and charges during the DCCI process. 
The solid orange lines represent solely the contribution from the core fluid with a high energy density ($e \geq e_{\mathrm{sw}}$) prior to particlization~\ref{active-core}. 


Figure~\ref{fig:conservation} (a) shows the time evolution of the energy during the time interval $\tau_0 \leq \tau < \tau_s$, where fluid formation by the source terms dominates over the cooling due to hydrodynamic expansion. 
Indeed, during this stage, most of the fluidized matter remains in the hot core ($e \geq e_{\mathrm{sw}}$) prior to particlization.
As discussed in Sec.~\ref{sec:color_string}, a slight violation of energy conservation occurs at $\tau \geq \tau_s$ due to the string modification processes, which remains to be improved in future work.


Figures~\ref{fig:conservation} (b), (c), and (d) show the time evolution of the baryon number, electric charge, and strangeness, respectively, for $\tau \geq \tau_s$. 
The remarkable behavior is that the baryon number and electric charge carried by the fluid (solid red) rapidly increase as a result of contributions from charge depositions; in other words, rapid equilibration of charges occurs. 
As the system evolves further, rapid cooling driven by hydrodynamic expansion takes over, and it is clearly seen that starting from around $\tau \approx 1$ fm/$c$, the baryon number and electric charge of the core progressively drop below $e_{\mathrm{sw}}$ and shift to the particlized phase. 
Due to strangeness neutrality, the average strangeness is expected to vanish, and only fluctuations around $S=0$ remain.
However, in this particular event, the strangeness in the core is predominantly positive at early times as a result of event-by-event fluctuations. 

In summary, the energy is exactly conserved for $\tau_0 \leq \tau < \tau_s$, while a slight violation occurs at later times, as discussed above.
In contrast, the conserved charges are conserved throughout the entire evolution.

\subsection{Profiles of thermodynamic variables}\label{sec:profiles}
One of the key features of the DCCI3 framework is that it describes how event-by-event charge density fluctuations in the initially produced partons are naturally carried over into the formation of a locally equilibrated fluid through dynamical initialization. 
In this section, we examine the resulting core profiles to illustrate how this feature manifests itself.

\begin{figure*}[htbp]
\includegraphics[width=0.75\linewidth]{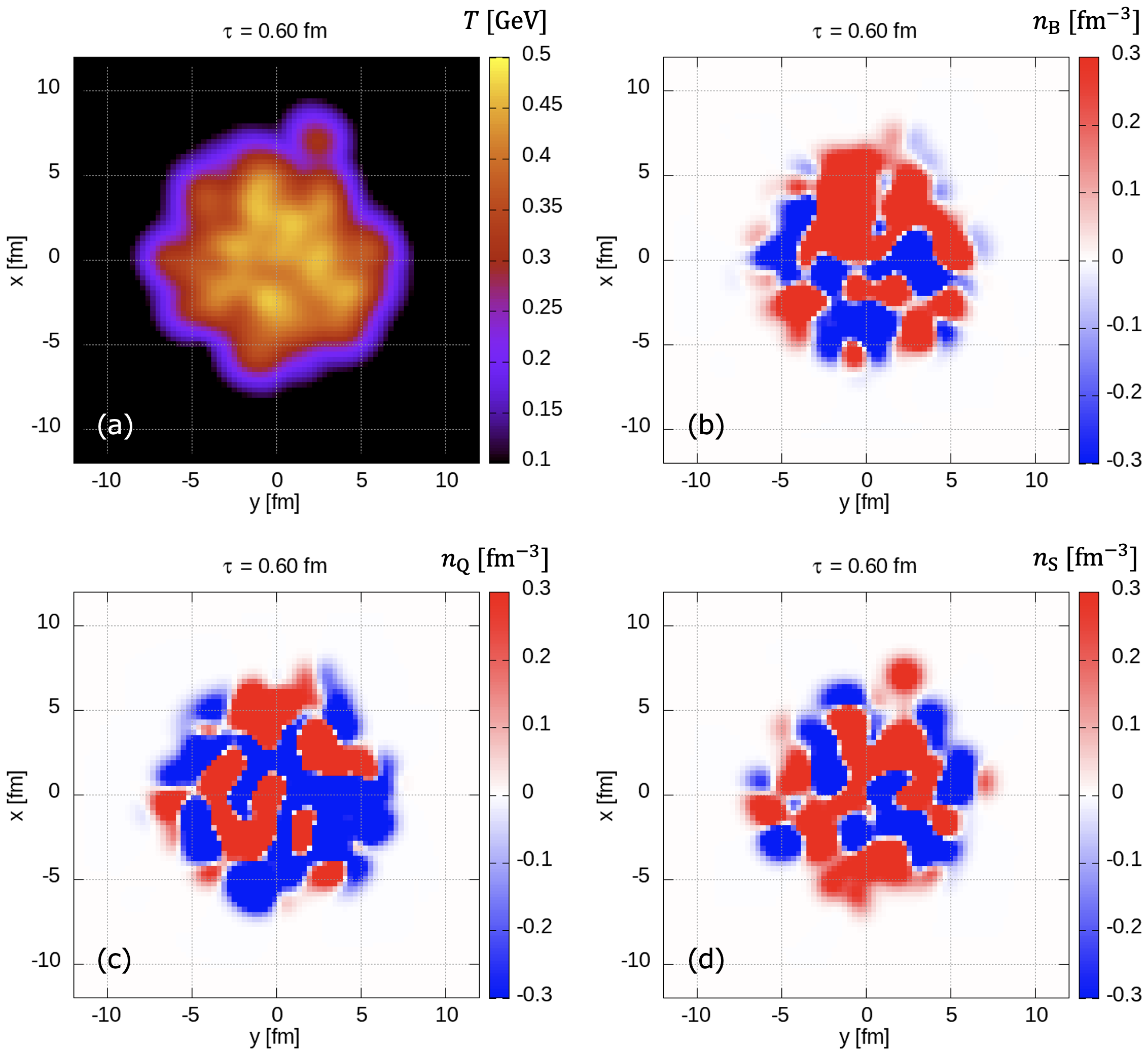}
\caption{Snapshots of thermodynamic variables of the core at $\tau = 0.6\,\mathrm{fm}$ in the transverse ($x$--$y$) plane at $\eta_s = 0$. Panels show (a) temperature, (b) baryon number density, (c) electric charge density, and (d) strangeness density. In each panel, the color bar indicates the magnitude of the variables, in units of $\mathrm{GeV}$ for temperature and $\mathrm{fm^{-3}}$ for charge densities.}
\label{fig:profile_tr}
\end{figure*}
Figure~\ref{fig:profile_tr} shows snapshots of thermodynamic variables of the core fluid at $\tau = 0.6\,\mathrm{fm}$ in the transverse ($x$–$y$) plane at $\eta_s = 0$ for a single event. 
As shown in panel (a), the temperature at the center of the core already reaches almost $0.5\,\mathrm{GeV}$ at this early time, $\tau = 0.6\,\mathrm{fm}$, as a consequence of rapid fluidization and heating.
Profiles of conserved charge densities are presented in panels (b), (c), and (d).
A notable feature is the appearance of localized regions with negative charge densities caused by individual parton depositions, visible as blue gradations.
Moreover, the absolute value of all charge densities reaches $0.3\,\mathrm{fm^{-3}}$, which corresponds to roughly twice the normal nuclear density.
While the net charge densities at midrapidity are globally expected to average to zero in high-energy collisions, these results demonstrate the presence of remarkably strong local fluctuations. 
These strong local fluctuations naturally arise because the discrete charge depositions from individual partons are explicitly taken into account, breaking the conventional assumption of homogeneously vanishing charge densities at midrapidity. 
Understanding how the local charge concentrations from these fluctuations influence the resulting chemical potentials and, consequently, final-state hadron production is an important question to be explored.

\begin{figure*}[htbp]
\includegraphics[width=0.75\linewidth]{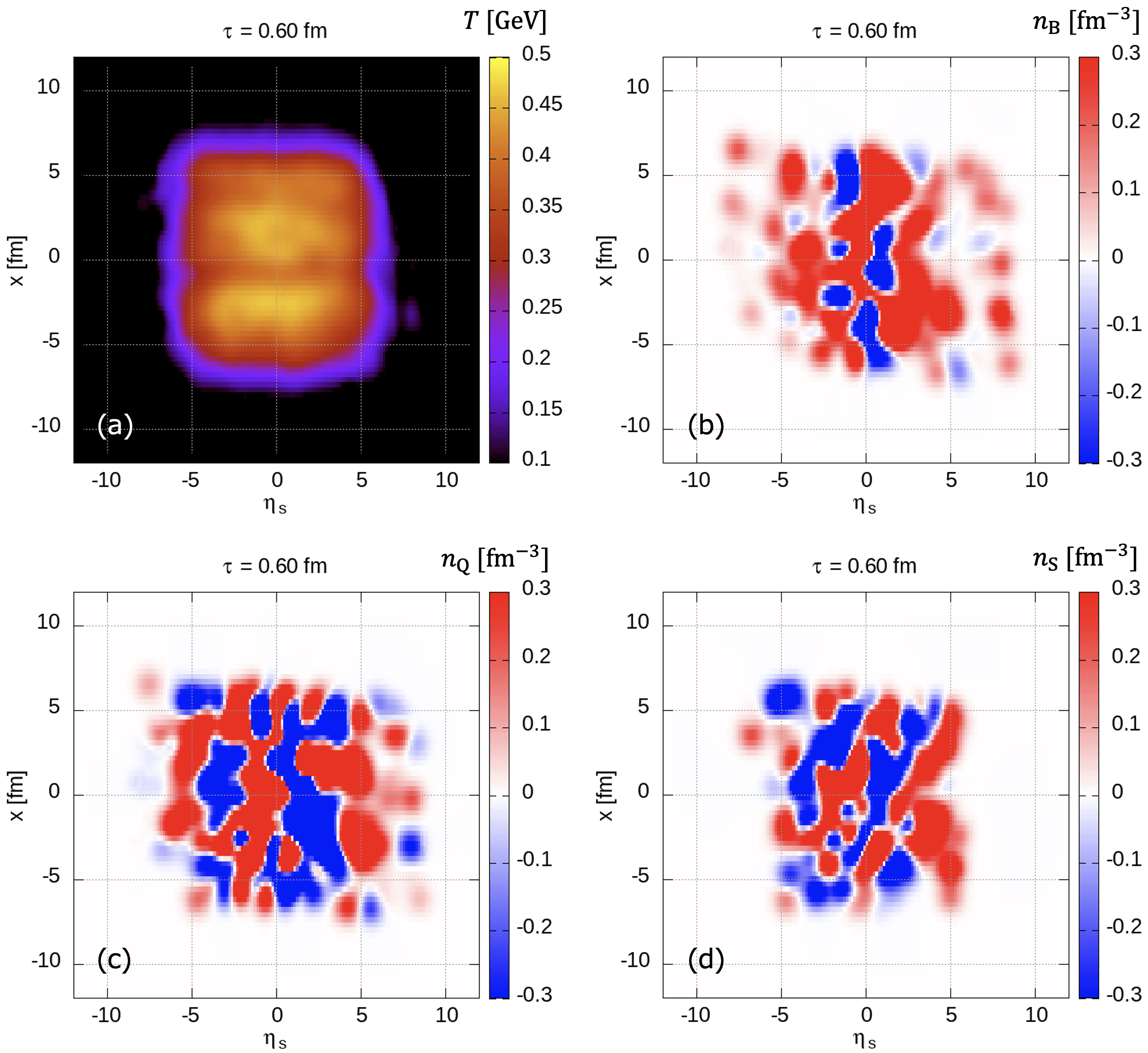}
\caption{Snapshots of thermodynamic variables of the core at $\tau = 0.6\,\mathrm{fm}$ in the longitudinal ($x$--$\eta_s$) plane at $y = 0$. Panels show (a) temperature, (b) baryon number density, (c) electric charge density, and (d) strangeness density. In each panel, the color bar indicates the magnitude of the variables, in units of $\mathrm{GeV}$ for temperature and $\mathrm{fm^{-3}}$ for charge densities.}
\label{fig:profile_long}
\end{figure*}
More interesting phenomena are seen in the longitudinal profiles (Fig.~\ref{fig:profile_long}).
As shown in panel (a), the medium fluid is formed widely along the space-time rapidity, with the temperature gradually decreasing toward the forward region. 
By examining the balance between positive and negative regions in the baryon number density [Fig.~\ref{fig:profile_long} (b)], 
they exhibit comparable magnitudes 
around midrapidity.
Turning to the forward region, the balance between positive and negative densities breaks down, and a net positive value becomes dominant, reflecting the influence of the initial baryon number carried by the projectile nuclei. 
Interestingly, as the overall amplitude of the fluid density decreases toward the forward region, $n_B$ becomes somewhat smaller, making it actually weaker than the local positive concentrations at midrapidity driven by the fluctuations. 
It should be emphasized that the baryon number density shown here represents only the equilibrated baryon number density, rather than the total baryon number of the system.
This is in contrast to conventional hydrodynamic models, in which the entire baryon number is typically assumed to be initialized in the fluid.
In this respect, it is remarkable that a non-negligible amount of baryon number is dynamically equilibrated in the forward region.


Similar features are also observed for the electric charge density [Fig.~\ref{fig:profile_long} (c)], which is broadly distributed along the longitudinal direction. 
It is important to note, however, that the dominance of the net positive charge in the forward region is less pronounced compared to the case of the baryon number. 
While the projectile nuclei possess a positive net charge as a whole, their initial valence distribution already contains local fluctuations of both signs at the partonic level. 
This stands in contrast to the baryon number, where all valence quarks possess a positive baryon number, leading to a more uniform dominance of the net positive density in the forward region.

Regarding the strangeness density [Fig.~\ref{fig:profile_long}(d)], its  magnitude is notably smaller in the forward rapidity region ($|\eta_s| \gtrsim 5$) compared to those of the baryon number and electric charge. 
Unlike baryon and electric charges, which are carried by the projectile nuclei, strange quarks are predominantly produced via $\mathrm{s}$--$\mathrm{\bar{s}}$ pair creation primarily around midrapidity. 
Consequently, the net strangeness density distribution does not extend as far longitudinally as the other conserved charges. 

\subsection{Chemical potentials on switching hypersurface}\label{sec:chemical_potentials}
Following the hydrodynamic space-time evolution, whose behavior was explored in the previous section  (Sec.~\ref{sec:profiles}), hadron sampling is performed on the switching hypersurface where the energy density falls below $e_{\mathrm{sw}}$. 
The local thermodynamic variables on this surface, such as temperature and chemical potentials, determine the momentum spectra and chemical composition of the sampled hadron species. 
In this section, we present the distributions of these thermodynamic variables along the space-time rapidity at the moment of switching. 
The results are averaged over the switching hypersurface elements across minimum-bias events.

\subsubsection{Averaged chemical potentials}
\begin{figure}[htbp]
\includegraphics[width=0.95\linewidth]{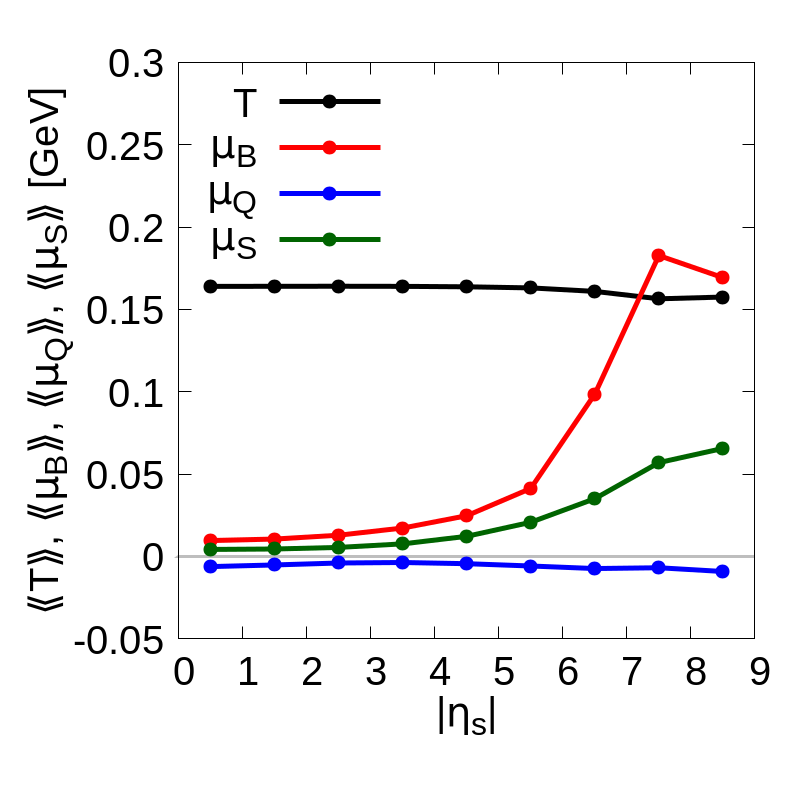}
\caption{Space-time rapidity dependence of averaged temperature and chemical potentials on switching hypersurface. Each line shows the temperature (black), baryon chemical potential (red), electric charge chemical potential (blue), and strangeness chemical potential (green). The results are averaged over hypersurface elements and minimum-bias events.}
\label{fig:T_mu}
\end{figure}
\begin{figure}[htbp]
\includegraphics[width=0.95\linewidth]{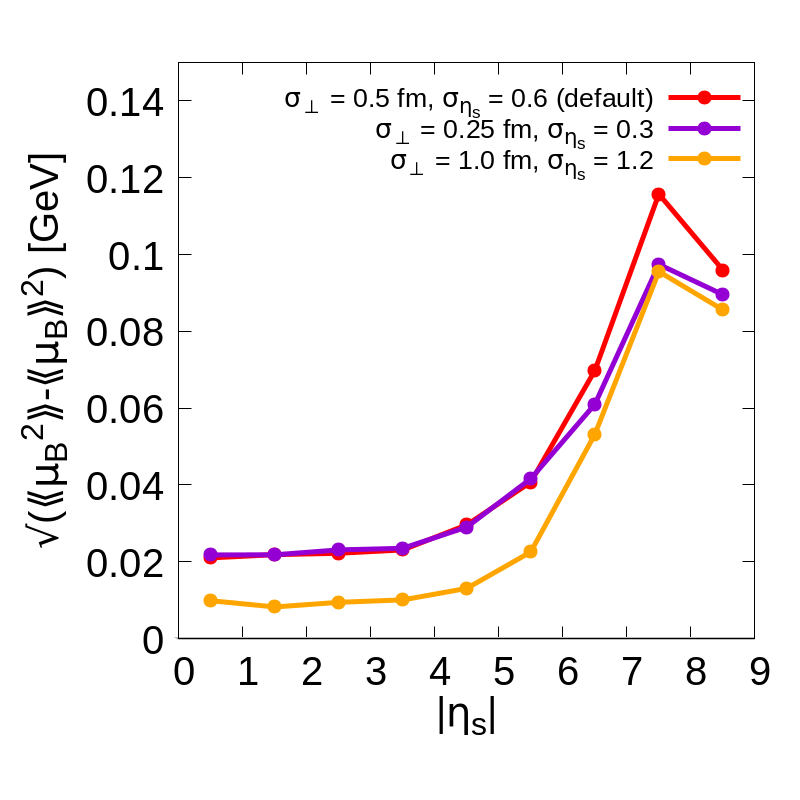}
\caption{Standard deviation of baryon chemical potential as a function of space-time rapidity. Lines show the default width ($\sigma_\perp = 0.5~\mathrm{fm}$ and $\sigma_{\eta_s} = 0.6$; red), half the default width ($\sigma_\perp = 0.25~\mathrm{fm}$ and $\sigma_{\eta_s} = 0.3$; purple), and twice the default width ($\sigma_\perp = 1.0~\mathrm{fm}$ and $\sigma_{\eta_s} = 1.2$; orange) of the Gaussian function in the conserved charge source terms.}
\label{fig:muB_meansquare}
\end{figure}

Figure~\ref{fig:T_mu} shows the space-time rapidity dependence of the temperature and chemical potentials. 
The baryon chemical potential (red) approaches zero at midrapidity, whereas it rises to approximately $200\,\mathrm{MeV}$ in the forward region, reaching a magnitude comparable to the temperature. 
This value aligns with the one realized in Au+Au collisions at $\sqrt{s_{NN}} = 19.6\,\mathrm{GeV}$ in the BES program at RHIC~\cite{STAR:2017sal}, demonstrating that the rapidity scan provides a complementary and effective approach to the BES program for exploring the QCD phase diagram.

The value of the baryon chemical potential at midrapidity, although it appears remarkably small in the figure, is calculated to be approximately $10\,\mathrm{MeV}$. 
This value is somewhat larger than the $\mu_{B} = 0.7 \pm 3.8\,\mathrm{MeV}$ extracted from the thermal fits at the LHC~\cite{Andronic:2017pug}. 
This discrepancy stems from the fact that the current PYTHIA8 configuration generates excessive baryon stopping, which leaves a residual positive net-baryon density even in the midrapidity region.


The electric charge chemical potential (blue) exhibits an almost uniform behavior along space-time rapidity with a slightly negative value. 
In terms of absolute magnitude, $\mu_{Q}$ is about an order of magnitude smaller than $\mu_{B}$, and its impact on particle production is expected to be very small. 
The slightly negative $\mu_Q$ observed here is qualitatively consistent with the value extracted in Pb+Pb collisions at $\sqrt{s_{NN}} = 5.02~\mathrm{TeV}$, $\mu_Q = -0.18 \pm 0.9~\mathrm{MeV}$~\cite{ALICE:2023ulv}, although a quantitative comparison is not intended due to the different collision energies.

It might appear counterintuitive that $\mu_{Q}$ takes a negative value while $\mu_{B}$ is positive. This behavior, however, can be understood as a competition between $\mu_{B}$ and $\mu_{Q}$ in determining the system's net charge. Even in the absence of $\mu_{Q}$, a positive $\mu_{B}$ inherently generates a significant positive charge by promoting the production of protons and neutrons in equal proportions ($Q/B \approx 0.5$). 
Since heavy nuclei, such as lead, are neutron-rich ($Q/B \approx 0.4$), the charge baseline produced by $\mu_{B}$ alone is already excessive. Consequently, a negative $\mu_{Q}$ must emerge to counteract this surplus, adjusting the net charge downward to match the composition of the colliding nuclei.


The strangeness chemical potential, $\mu_{S}$ (green), exhibits an increasing trend toward the forward region. 
Notably, near the beam rapidity ($Y_{\mathrm{beam}} \approx 8$), it reaches approximately $70\,\mathrm{MeV}$, suggesting a non-negligible impact on strangeness production in this region. 
This rise is driven by the requirement of strangeness neutrality. 
As $\mu_{B}$ increases, the production of strange baryons—which carry $\mathrm{s}$-quarks—is also enhanced naturally, shifting the net strangeness of the medium toward a negative value ($S = -1$ per $\mathrm{s}$-quark). To counteract this and restore strangeness neutrality, a positive $\mu_{S}$ is required to facilitate the production of $\bar{\mathrm{s}}$-quarks over $\mathrm{s}$-quarks.

\subsubsection{Fluctuations of baryon chemical potential}
\begin{figure*}[t]
\includegraphics[width=0.95\linewidth]{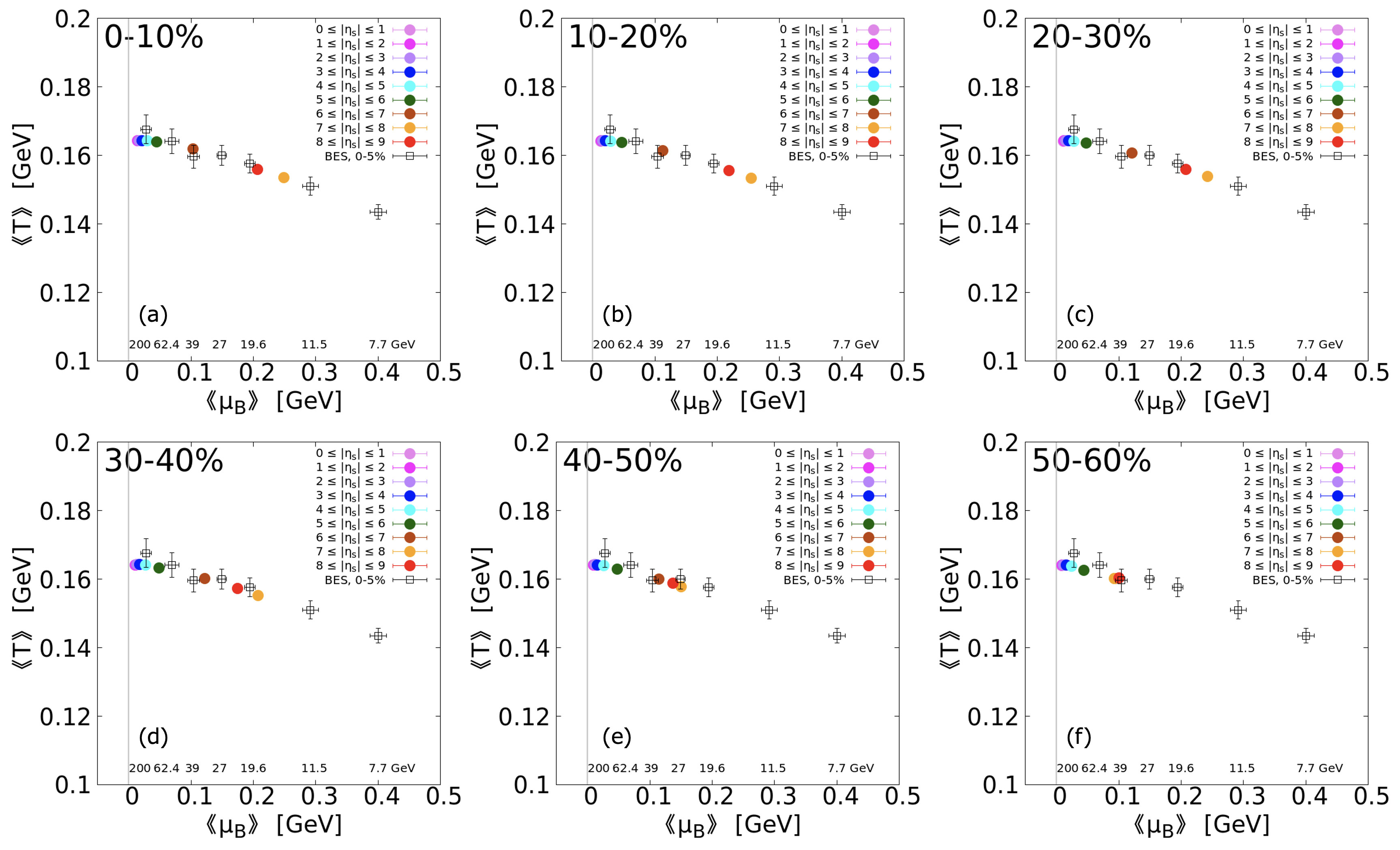}
\caption{Centrality dependence of averaged temperature and baryon chemical potential on $T$--$\mu_{B}$ plane. Events are separated into six centrality classes: (a) 0-10\%, (b) 10-20\%, (c) 20-30\%, (d) 30-40\%, (e) 40-50\%, and  (f) 50-60\%. The colored markers in each panel denote the averaged temperature and baryon chemical potentials evaluated within specific rapidity intervals. For reference, the baryon chemical potentials extracted from yield fits to BES data~\cite{STAR:2017sal} (0-5\%) are also shown as open squares.}
\label{fig:centrality}
\end{figure*}
Next, the typical magnitude of the fluctuations of $\mu_{B}$ is examined. 
The finite standard deviation arises precisely because the dynamical initialization procedure leads to the emergence of localized deposition of conserved charges. 
Figure~\ref{fig:muB_meansquare} shows the standard deviation of the baryon chemical potential as a function of space-time rapidity for various values of the Gaussian smearing widths, $\sigma_\perp$ and $\sigma_{\eta_s}$, appearing in the source terms of Eq.~\eqref{eq:charge_source}. 
These smearing widths are key parameters in the DCCI3 model that decide the size of the local concentrations of baryon-number density formed by the charge deposition, thereby controlling the resulting fluctuations in the baryon chemical potential. 

It is found that, for the default (red) and half-width (purple) smearing cases, the typical fluctuation of $\mu_{B}$ reaches approximately $20~\mathrm{MeV}$ even at midrapidity. 
This value is substantially larger than the baryon chemical potential extracted from yield fits at the LHC, $\mu_{B} = 0.7 \pm 3.8~\mathrm{MeV}$~\cite{Andronic:2017pug}. 
Even in the larger smearing case (orange), the typical fluctuation remains approximately $10~\mathrm{MeV}$. 
This result indicates that the conventional assumption of vanishing chemical potential at the LHC should be reconsidered when pursuing a more precise description of QCD matter. 
Toward forward rapidities, the fluctuations become even larger, with typical variations of $\mu_{B}$ reaching on the order of $100~\mathrm{MeV}$.
This behavior can be understood from the fact that the magnitude of the $\mu_{B}$ is largely controlled by the ratio $n_{B}/s$.
In the forward region, the entropy density $s$ is smaller than at midrapidity. 
As a result, even a modest baryon number deposition leads to a relatively large change in $n_{B}/s$, making $\mu_{B}$ more sensitive to baryon number fluctuations and thereby enhancing its event-by-event variation.

\subsubsection{Centrality dependence}
The amount of baryon number deposited in the reaction region is expected to depend on the collision geometry, i.e., the centrality of the collision. 
Consequently, the behavior of the $\mu_{B}$ can also vary among different centrality classes. 
Figure~\ref{fig:centrality} shows the centrality dependence of averaged temperature and baryon chemical potential on $T$--$\mu_{B}$ plane. 
The colored markers show $\langle\langle T \rangle\rangle$ and $\langle\langle \mu_B \rangle\rangle$ at the moment of particlization in Pb+Pb collisions at $\sqrt{s_{NN}} = 2.76$~TeV from DCCI3, averaged over fluid cells for each rapidity window $|\eta_s|$.
Since $T$ and $\mu_B$ of each fluid cell at particlization are determined by the switching condition $e = e_\mathrm{sw}$ on the equation of state (NEOS-4D), the results lie on the corresponding switching hypersurface projected onto the $T$--$\mu_B$ plane. 
The open markers show $T_\mathrm{ch}$ and $\mu_B$ at chemical freeze-out obtained by fitting the THERMUS statistical thermal model to experimental particle yield ratios measured in Au+Au collisions at different collision energies from the STAR BES program data~\cite{STAR:2017sal}. 

One can see that, for all centrality classes, the DCCI3 results for different rapidity windows align with the BES results at different collision energies.
The rapidity-scan coverage demonstrated by DCCI3 varies non-monotonically with centrality: while the coverage generally shifts toward smaller $\langle\langle \mu_B \rangle\rangle$ values from central to peripheral collisions---reflecting the fact that peripheral collisions involve weaker baryon stopping and produce less fluid---the largest $\langle\langle \mu_B \rangle\rangle$ is reached not in the 0--10\% class but in the 10--20\% class, where the forward rapidity windows extend the coverage up to $\langle\langle \mu_B \rangle\rangle \sim 0.25$~GeV, almost corresponding to $\sqrt{s_{NN}} = 11.5~\mathrm{GeV}$ collisions in the BES program. 
This non-monotonic behavior can be understood as arising from the competition between baryon stopping and multi-parton interactions (MPI). While both baryon stopping and MPI are stronger in more central collisions, MPI increases the entropy density and thus suppresses $n_B/s$, which roughly controls the magnitude of $\mu_B$. In the 0--10\% centrality class, the large MPI activity dominates, leading to a suppression of $\mu_B$. In the 10--20\% class, MPI is slightly weaker, and the reduction in entropy density outweighs the reduction in baryon stopping, resulting in the largest $\langle\langle \mu_B \rangle\rangle$ in the forward rapidity region.

\subsection{Net particle yields}\label{sec:particle_yields}
\begin{figure}[htbp]
\includegraphics[width=0.95\linewidth]{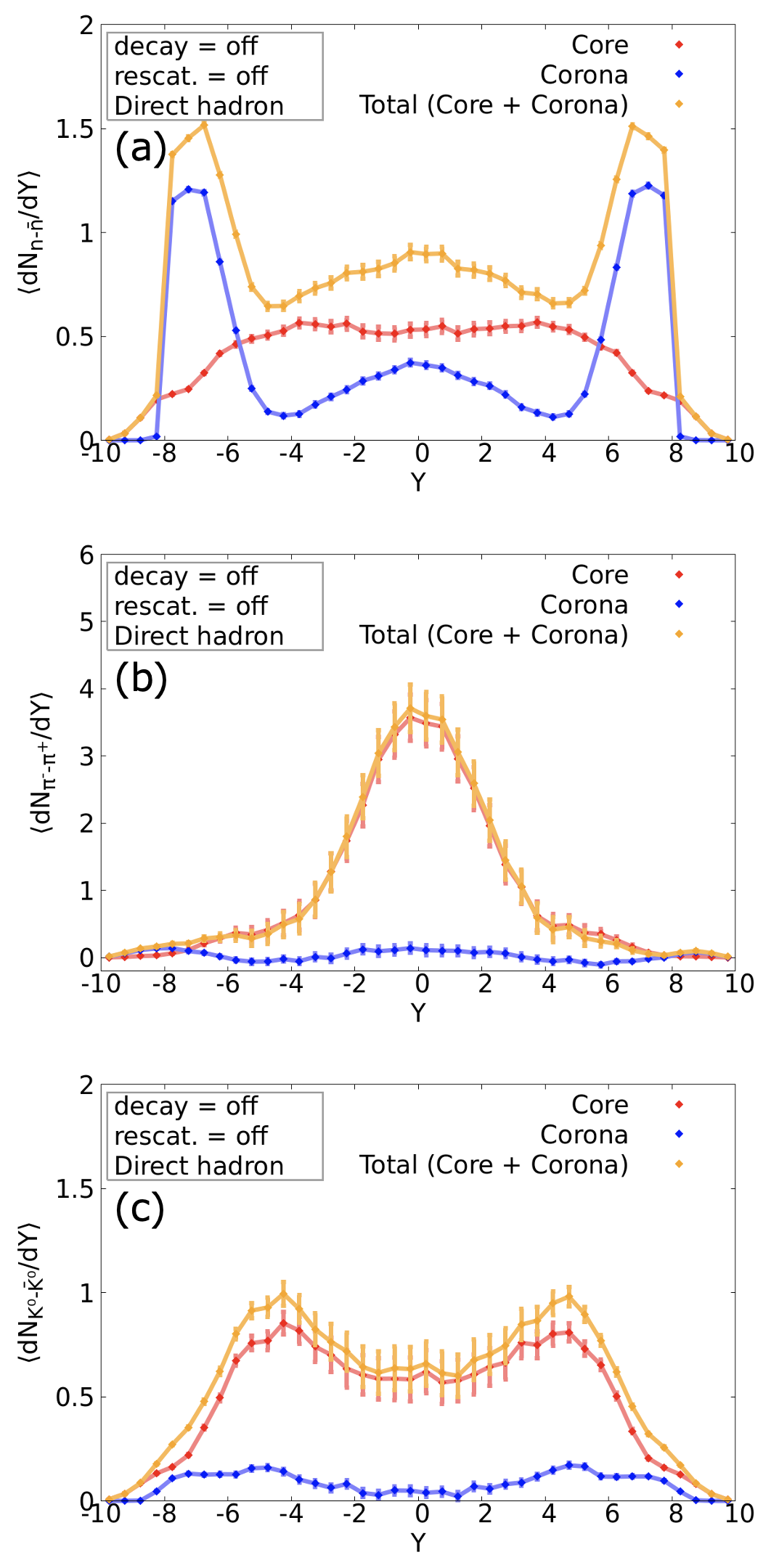}
\caption{Rapidity distributions of net hadrons directly obtained from hadronization of the core and the corona. Panels show (a) ${\mathrm{n} - \mathrm{\bar{n}}}$, (b) $\pi^- - \pi^+$, and (c) $\mathrm{K^0 - \bar{K}^0}$. The results are decomposed into core-origin (red) and corona-origin (blue), with their sum shown by the orange line. Both hadronic rescatterings and decays are switched off to display the direct yields from the core and the corona.}
\label{fig:effects_of_mu}
\end{figure}
The extension of the DCCI framework to finite conserved charges enables us to obtain locally varying chemical potentials on the switching hypersurface, which are subsequently reflected in hadron production through Cooper--Frye sampling implemented in iS3D.
In this subsection, we examine the effects of each chemical potential on hadron production by disabling hadronic decays and rescatterings of JAM, so that hadrons from the core directly reflect its thermodynamic variables. 

\begin{figure*}
\includegraphics[width=0.95\linewidth]{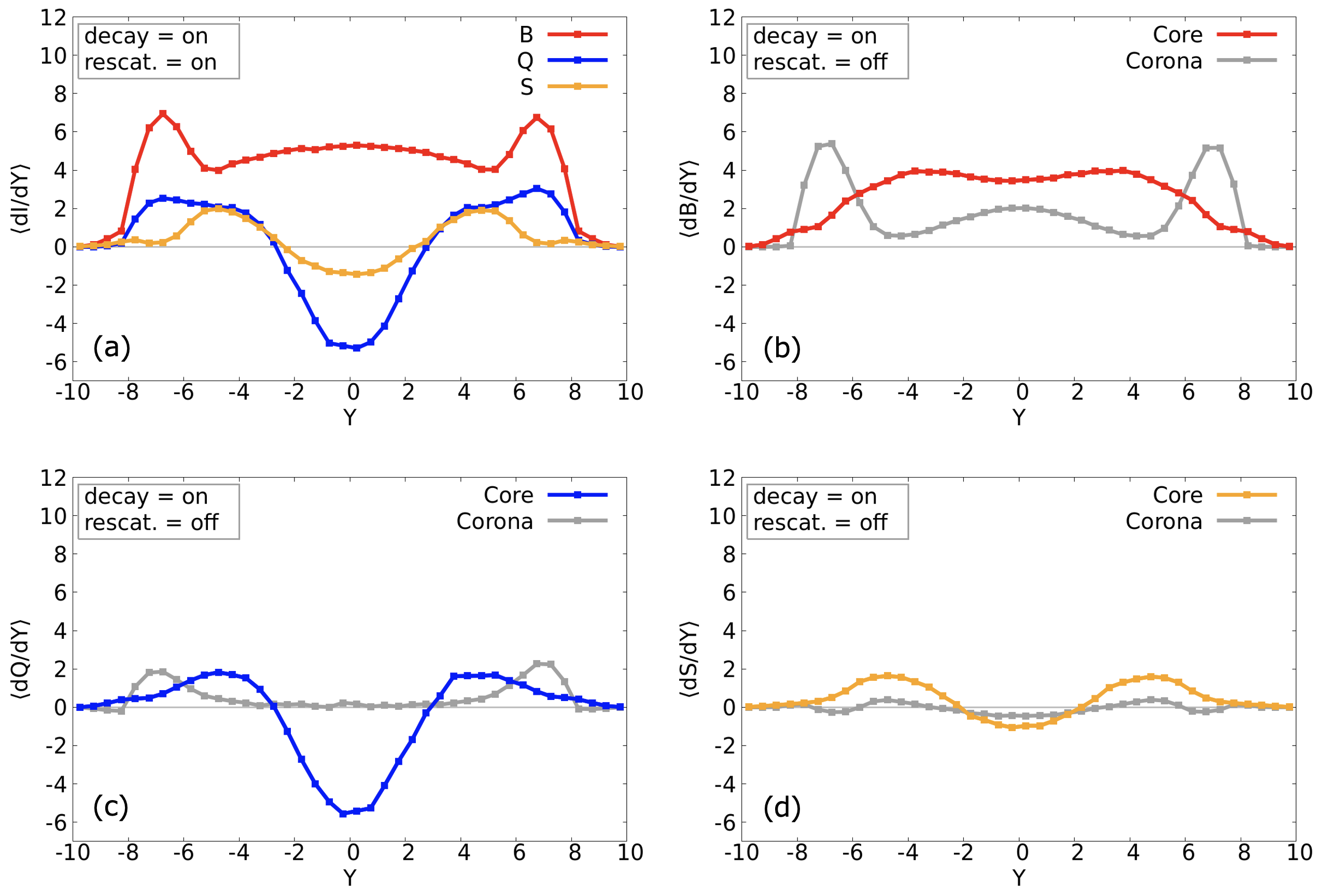}
\caption{Rapidity distributions of conserved charges carried by final-state hadrons. In panel (a), all conserved charge distributions are shown, including rescatterings and decays. Panels (b), (c), and (d) present each conserved charge with a core-corona separation. Colored lines denote the core contribution, while gray lines represent the corona contribution. In panels (b), (c), and (d), hadronic rescatterings are switched off while decays are retained to enable the core-corona separation.}
\label{fig:charge_dist}
\end{figure*}


Figure~\ref{fig:effects_of_mu} (a) shows the rapidity distribution of net neutrons ($
{\mathrm{n} - \mathrm{\bar{n}}}$).
Since neutrons carry the quantum numbers ($B=1$, $Q=0$, $S=0$) and antineutrons carry ($B=-1$, $Q=0$, $S=0$), 
the net neutron yield from the core directly reflects the baryon chemical potential on the switching hypersurface.
Examining the core (red), one observes a broad enhancement of the net neutron yield across a wide rapidity region, with a gentle and wide peak around midrapidity that falls off toward forward rapidity, where the volume of the locally equilibrated core is significantly smaller. 
On the other hand, an enhancement of net neutrons is also observed in the nonequilibrium corona (blue), particularly in the vicinity of the beam rapidity.

In Fig.~\ref{fig:effects_of_mu} (b), the rapidity distribution of net negatively charged pions ($\pi^{-}-\pi^{+}$) is shown.
The negatively charged pions carry $(B=0,\, Q=-1,\, S=0)$ and positively charged pions carry $(B=0,\, Q=+1,\, S=0)$.
Namely, the net negatively charged pion yield is entirely determined by the electric charge chemical potential. 
%
Here, we take $\pi^{-}-\pi^{+}$ rather than $\pi^{+}-\pi^{-}$ to keep the yield positive overall, since the averaged $\mu_Q$ is negative as shown in Fig.~\ref{fig:T_mu}. 
Although $\mu_Q$ is nearly uniform in rapidity, the $\pi^{-} - \pi^{+}$ distribution is concentrated at midrapidity. 
It falls off in the forward region, reflecting the smaller core volume there. 
As discussed later, these net negatively charged pions significantly affect the charge distributions of final-state hadrons.
In contrast, the corona contribution to the net yield of negatively charged pions remains close to zero.

Figure~\ref{fig:effects_of_mu} (c) shows the results for the net neutral kaons ($\rm{K^{0} - \bar{K}^{0}}$).
Since $\rm{K^{0}}$ carries the quantum numbers $(B=0,\, Q=0,\, S=1)$, while $\rm{\bar{K}^{0}}$ carries $(B=0,\, Q=0,\, S=-1)$, this observable is well suited for probing the influence of the strangeness chemical potential.
The core contribution is enhanced and reaches a maximum around $|Y| \approx 5$.
While $\mu_B$ and $\mu_S$ exhibit qualitatively similar rapidity dependences, the resulting net-hadron distributions need not show the same structure. The presence and location of a peak are determined by the interplay between the decreasing core volume and the chemical-potential-induced enhancement of the net yield, whose relative importance depends on both the magnitude of the relevant chemical potential and the hadron mass.
In contrast to the core distribution, the corona contribution remains nearly zero, similar to the case of net charged pions.

\subsection{Charge distribution}\label{sec:charge_distribution}
In this section, we present the rapidity distributions of the conserved charges ($B$, $Q$, and $S$) carried by final-state hadrons produced in Pb+Pb collisions at $\sqrt{s_{NN}} = 2.76~\rm{TeV}$. 
Using DCCI3, we can trace these charges from midrapidity to the forward region. 
At the same time, we can separately identify which portions of the charges come from equilibrated matter (core) and which from non-equilibrated partons (corona) by switching off hadronic rescatterings in the simulations.
This dual information reveals the three-dimensional structure of conserved charges throughout the collision system. 

The rapidity distribution of final-state conserved charges from full DCCI3 simulations with both rescatterings and decays enabled in JAM is shown in Fig.~\ref{fig:charge_dist} (a). 
Focusing first on the baryon number (red), one finds two prominent peaks near the beam rapidities originating from the leading baryons, followed by another peak developing toward midrapidity.
The latter can be attributed to the formation of equilibrated matter with finite $\mu_{B}$ and a large fluidized volume in this region.

Turning to the electric charge (blue) in panel (a), a naive expectation based on the property of lead nuclei ($Q/B \approx 0.4$) would suggest a distribution similar in shape to that of the baryon number.
However, the results show a qualitatively different behavior.
As demonstrated in Fig.~\ref{fig:T_mu}, the average $\mu_{Q}$ of the system is entirely negative.
This leads to an imbalance in the charged pion yields, particularly around midrapidity, as seen in Fig.~\ref{fig:effects_of_mu} (b).
Consequently, a negative dip structure develops around midrapidity in the charge distribution.
Toward forward rapidity, where the amount of equilibrated matter is expected to be small, and the influence of chemical potentials is therefore reduced, the expected relation $Q/B \approx 0.4$ is approximately recovered.

The most intriguing behavior appears in the strangeness distribution (orange) in panel (a).
Although the initial strangeness is nearly zero across all rapidities due to strangeness neutrality, the final distribution develops peak structures around $|Y|\approx 5$.
This feature is strongly correlated with the behavior of $\rm{K^{0} - \bar{K}^{0}}$ shown in Fig.~\ref{fig:effects_of_mu} (c), which reflects the enhancement of $\mu_{{S}}$ in the forward region.
At midrapidity, however, the final net strangeness becomes slightly negative, a behavior that cannot be explained solely by $\mu_{{S}}$.
To understand this, for example, one must consider the lightest strange baryon, the $\Lambda$ particle.
Since $\Lambda$ baryons are influenced by both $\mu_{{B}}$ and $\mu_{{S}}$, and $\mu_{{B}} > \mu_{{S}}$ in the equilibrated matter, their yields are predominantly affected by $\mu_{{B}}$.
As a result, yields exhibit $\Lambda > \bar{\Lambda}$, causing the net strangeness to become negative.
This demonstrates that the strangeness distribution exhibits a non-trivial imbalance beyond simple strangeness neutrality, implying complications in the interpretation of strangeness-related observables, which will be discussed in a forthcoming paper.

Panels (b), (c), and (d) depict the core-corona separated distributions of $B$, $Q$, and $S$, respectively.
The baryon number distribution shown in Fig.~\ref{fig:charge_dist} (b) closely resembles the net neutron distributions in Fig.~\ref{fig:effects_of_mu} (a) for both the core and the corona, indicating that $\mu_{{B}}$ has a substantial impact on the final-state baryon distribution.
For the electric charge shown in panel (c), the non-trivial structure observed in panel (a) is found to originate predominantly from the core component, indicating that it is driven by the $\mu_{{Q}}$.
In contrast, the electric charge distribution in the corona closely follows the baryon number distribution in the corona, suggesting that the ratio $Q/B \approx 0.4$ is approximately maintained in the nonequilibrium component.
A similar conclusion holds for strangeness in panel (d): the non-trivial features in the distribution arise mainly from the core component, suggesting that $\mu_{S}$ also plays a non-negligible role in the final-state observables.
In contrast, the strangeness distribution in the corona component remains close to zero, indicating that strangeness neutrality is approximately maintained in the nonequilibrium component.

Although the conserved charge distributions shown here correspond to final-state hadrons, one should keep in mind that fluctuations accompany the hadronization procedure; therefore, the plotted distributions do not reflect exact conservation.
In addition, limitations in the validity range of the NEOS-4D equation of state, as well as uncertainties associated with the matching to the hadron resonance gas, can affect the conservation laws~\cite{Monnai:2024pvy}.
In fact, during the hydrodynamic evolution, a non-negligible number of fluid cells reach the upper boundary of the tabulated equation of state, which may introduce additional systematic effects.
Extending the range of the equation of state is therefore highly desirable for future studies.
Nevertheless, it should be emphasized that the DCCI module itself strictly enforces the conservation of all conserved charges, as demonstrated in Fig.~\ref{fig:conservation}.

\section{SUMMARY}\label{sec:summary}
In this study, we have developed a new dynamical framework, DCCI3, which extends the dynamical core–corona initialization model by incorporating the full dynamics of conserved charges, namely baryon number, electric charge, and strangeness, within a unified (3+1)-dimensional hydrodynamic description, including updates of the equation of state and particlization procedure. 
In contrast to conventional approaches that primarily focus on energy--momentum dynamics and often assume vanishing conserved charges at high collision energies, the present framework consistently treats both energy--momentum and conserved charges through phenomenological source terms, ensuring their conservation throughout the entire evolution.

A key feature of DCCI3 is the dynamical description of the equilibration process. 
Rather than imposing a global initialization time, the model continuously converts nonequilibrium partons into an equilibrated fluid through microscopic interactions. 
This allows for a natural separation between the equilibrated core and the non-equilibrated corona, while simultaneously describing their mutual interactions. 
Importantly, conserved charges are also dynamically deposited into the fluid in a fully consistent manner, enabling the evaluation of local charge densities and corresponding chemical potentials across the entire space-time evolution.

Applying this framework to Pb+Pb collisions at $\sqrt{s_{NN}} = 2.76~\rm{TeV}$, we have systematically investigated the space-time evolution of conserved charges and their associated chemical potentials. 
One of the central findings is that sizable fluctuations of chemical potentials emerge even in the midrapidity region, where they are conventionally assumed to vanish. 
In particular, the baryon chemical potential around midrapidity exhibits event-by-event fluctuations reaching up to $\pm 20~ \rm{MeV}$, despite its small average value. 
This result clearly demonstrates that the commonly adopted assumption of vanishing chemical potentials at the LHC is an oversimplification when considering local properties of the system.
In fact, within a framework incorporating fluctuations of conserved charges ($B$, $Q$, and $S$), it has been shown that the flow of identified particles exhibits a non-negligible breaking of particle--antiparticle symmetry~\cite{Gardim:2024nyz}.
In particular, particles carrying larger conserved charges, e.g., $\Omega$ baryons, are more sensitive to BQS fluctuations, and higher-order flow coefficients such as $v_3$ are found to be more sensitive than $v_2$~\cite{Gardim:2024nyz}.
These results indicate that the dynamics of conserved charges play an essential role in enabling more detailed discussions of nuclear substructure, as well as in constructing precise backgrounds for the search for critical fluctuations.

Furthermore, toward forward rapidity, the baryon chemical potential increases rapidly and becomes comparable to the temperature. 
The accessible value of chemical potentials can be systematically controlled through rapidity intervals and centrality selections.
The present results therefore suggest that the Rapidity Scan (RS) can serve as a complementary approach to the Beam Energy Scan (BES) for exploring the QCD phase diagram, providing access to high-$\mu_{{B}}$ physics without lowering the collision energy.

The analysis of hadron production further reveals that these finite chemical potentials leave clear imprints on final-state observables. 
The non-trivial structures observed in the rapidity distributions of conserved charges are shown to originate predominantly from the core component, highlighting the essential role of charge dynamics in shaping final-state observables.

Another important outcome of this study is the demonstration that conserved charges are equilibrated over a wide space-time rapidity range, including forward regions where equilibration is not guaranteed so far. 
At the same time, the interplay between the magnitude of chemical potentials and the corresponding equilibrated volume plays a crucial role in determining the final distributions. 
This competition leads to non-trivial rapidity structures that cannot be captured in simplified approaches.

Overall, the DCCI3 framework provides one of the most comprehensive dynamical descriptions of high-energy nuclear collisions currently available, covering a wide range of physics from low to high transverse momentum, from small to large systems, and from midrapidity to forward rapidity within a single unified approach. 
By incorporating conserved charge dynamics explicitly, the model opens a new avenue for precision studies of QCD matter and offers a robust theoretical foundation for future investigations, particularly in the context of forward physics.

The present work establishes the theoretical framework and elucidates the dynamical behavior of conserved charges. 
In a forthcoming paper, we will further explore the quantitative impact of these dynamics on experimental observables, including predictions for future measurements.

\begin{acknowledgments}
We are grateful to K.~Kuroki, A.~Shigeyoshi, and N.~Ito for valuable discussions, and to Y.~Kanakubo for sharing the DCCI2 source code, which was indispensable for this study.
The work by S.F. was supported by JST SPRING Grant No.~JPMJSP2169. 
The work by Y.T. was supported by JSPS KAKENHI Grant No.~25K07303. 
The work by T.H. was partly supported by JSPS KAKENHI Grant No.~JP23K03395.
\end{acknowledgments}

\appendix
\section{Result from NEOS-BQS}\label{sec:NEOSBQS}
\begin{figure}[htbp]
\includegraphics[width=0.95\linewidth]{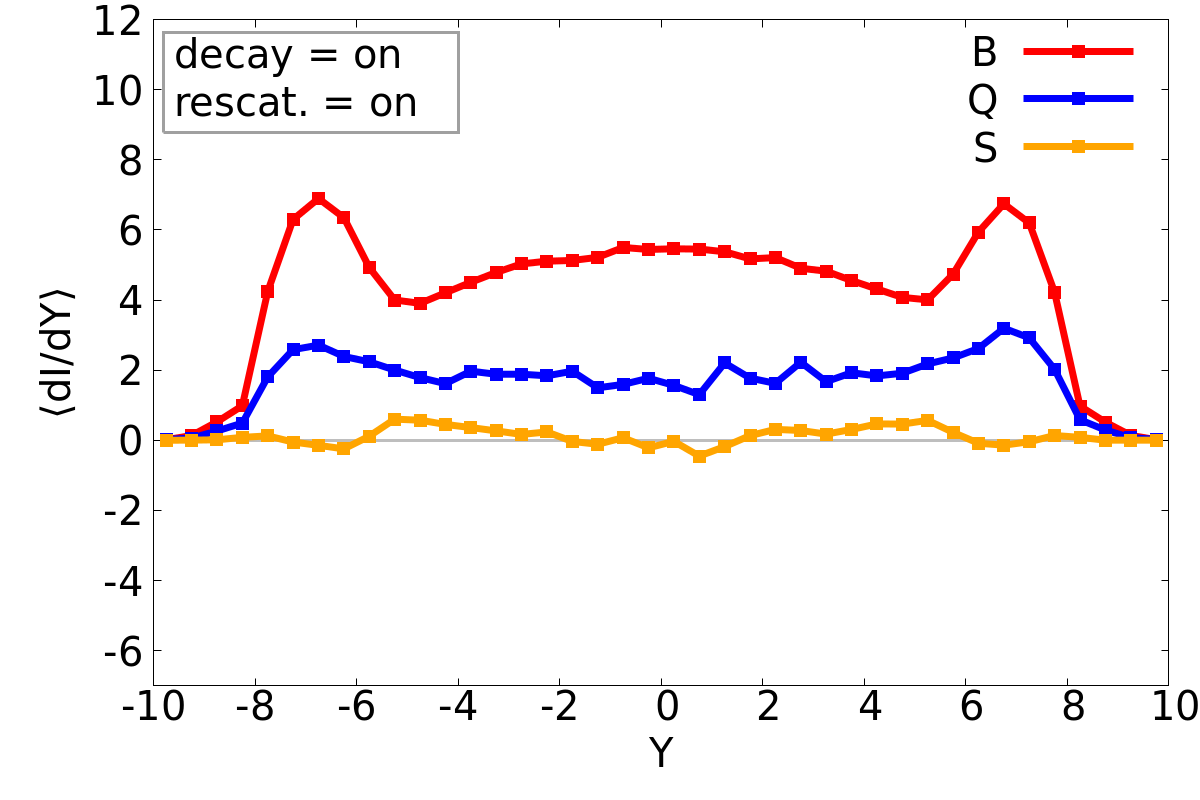}
\caption{Rapidity distributions of conserved charges carried by final-state hadrons, obtained using NEOS-BQS~\cite{Monnai:2019hkn, Monnai:2021kgu}. The red, blue, and orange curves represent baryon number, electric charge, and strangeness, respectively. The results include the effects of hadronic rescatterings and decays.}
\label{fig:charge_dist_neosbqs}
\end{figure}
In this study, the equation of state NEOS-4D~\cite{Monnai:2024pvy} is employed, which includes four independent degrees of freedom, $e$, $n_{{B}}$, $n_{{Q}}$, and $n_{{S}}$.
As a result, each conserved charge can evolve independently in spacetime.
Along with the implementation of NEOS-4D in DCCI3, the equation of state NEOS-BQS (-BS,~-B)~\cite{Monnai:2019hkn, Monnai:2021kgu} is also introduced for comparison.
NEOS-BQS is characterized by two independent variables, $e$ and $n_{{B}}$, while imposing constraints corresponding to the conditions in lead nuclei, namely a fixed ratio $n_{{Q}}/n_{{B}} = 0.4$ and strangeness neutrality $n_{{S}} = 0$.
By comparing the charge dynamics obtained with these two equations of state, the effects of fluctuations as well as deviations from the conditions in lead nuclei can be effectively isolated.

Figure~\ref{fig:charge_dist_neosbqs} shows the charge distributions of the final hadrons obtained using NEOS-BQS.
As a consequence of the constraints imposed in NEOS-BQS, the conserved charge distributions of the final-state hadrons are modified.
Although the baryon number distribution (red) appears quite similar to that obtained with NEOS-4D [Fig.~\ref{fig:charge_dist} (a)], notable differences appear in the electric charge (blue) and strangeness (orange) distributions.
It is evident that the distributions of electric charge and strangeness strictly follow the constraints imposed in NEOS-BQS.
Consequently, it is evident that the non-trivial structures observed in the $Q$ and $S$ distributions in Fig.~\ref{fig:charge_dist} (a) arise from conserved charge dynamics beyond these imposed assumptions.

\bibliography{apssamp}

\end{document}